\RequirePackage[OT1]{fontenc}
\documentclass[letterpaper,onecolumn,journal]{IEEEtran}
\usepackage{amsmath,amssymb,amsthm,mathtools,bm}

\usepackage{graphicx}
\usepackage{xcolor}
\usepackage{booktabs,longtable,tabularx,array,multirow}
\usepackage{float}

\usepackage{enumitem}
\usepackage{cite}
\usepackage[hidelinks]{hyperref}
\newtheorem{theorem}{Theorem}
\newtheorem{lemma}[theorem]{Lemma}
\newtheorem{proposition}[theorem]{Proposition}
\newtheorem{corollary}[theorem]{Corollary}

\theoremstyle{definition}
\newtheorem{definition}[theorem]{Definition}

\theoremstyle{remark}
\newtheorem{remark}[theorem]{Remark}

\newcommand{\llbracket}{[\![}
\newcommand{\rrbracket}{]\!]}

\newcolumntype{Y}{>{\raggedright\arraybackslash}X}

\hypersetup{
  pdftitle={Bounds for General (r,delta)-Quantum Locally Recoverable Codes}
}

\begin{document}

\title{Bounds for General $(r,\delta)$-Quantum
Locally Recoverable Codes}

\author{Shuowei~Ma and Fei~Shi%
\thanks{The authors are with the Institute of Quantum Computing
and Software, School of Computer Science and Engineering,
Sun Yat-sen University, Guangzhou 510006, China
(e-mail: shif26@mail.sysu.edu.cn).}%
\thanks{This work was supported in part by the National Natural Science Foundation of China (Grant Nos. 62601907 and 12571493) and by the 2026 Basic Start-up Fund of Sun Yat-sen University (Grant No. 67000-12266020).}%
}
\maketitle
\begin{abstract}
We study general $(r,\delta)$-quantum locally recoverable codes (qLRCs) without assuming purity or a stabilizer structure, and with arbitrarily overlapping recovery blocks. Using a selection lemma for multiple-erasure recovery, we establish a Singleton-like upper bound for general quantum subspaces. When $\delta=2$, this bound reduces to the general single-erasure bound of Golowich and Guruswami. An integer optimization yields an explicit remainder correction that tightens the bound by up to $\delta-2$. For pure codes, local decoupling and entropy arguments extend the parameter bound of Galindo et al., originally obtained for pure stabilizer codes constructed from Euclidean or Hermitian dual-containing classical codes, to arbitrary pure quantum subspaces. We also remove the partition assumption from several existing pure-code bounds for disjoint recovery blocks and obtain a further remainder correction by optimizing the budget. Finally, we analyze three families attaining these bounds: pure codes with overlapping recovery blocks and $d=\delta$, which exhibit both a strict improvement of the pure-code bound and simultaneous equality in the pure-code and general bounds; impure codes with disjoint recovery blocks attaining the general bound; and an existing family of pure codes with overlapping recovery blocks and $d>\delta$, which simultaneously attains all these bounds.
\end{abstract}
\begin{IEEEkeywords}
Quantum locally recoverable codes, Singleton-like bounds, Multiple-erasure recovery.
\end{IEEEkeywords}

\section{Introduction}\label{sec:introduction}
In large-scale distributed storage systems, locally recoverable codes (LRCs) reduce repair costs by limiting the number of other nodes accessed to repair a failed node~\cite{Huang2007Pyramid,Gopalan2012Locality}. Specifically, a code has locality $r$ if each code symbol is uniquely determined by at most $r$ other code symbols. This principle also applies to quantum storage, where recovery must additionally preserve the encoded state and its correlations with an external reference system. These requirements call for a corresponding theory of quantum local recovery.

Golowich and Guruswami introduced quantum locally recoverable codes
(qLRCs) and constructed quantum Tamo--Barg codes and their folded
variants~\cite{Golowich2023LRC}. Subsequent single-erasure constructions
use CSS, polynomial, Hermitian, and cyclic-code
methods~\cite{Luo2023CSS,Sharma2025Good,Xie2025Optimal,Li2026Hermitian,Yang2026Small}.
Galindo et al. introduced quantum $(r,\delta)$-local recovery and
characterized it for stabilizer codes~\cite{Galindo2026Locality}.
Further constructions employ matrix-product
methods~\cite{Galindo2026MPC,CaoZhou2026MPC,CaoZhou2026MP},
classical code decompositions~\cite{zhou2025optimalquantumrdeltalocallyrepairable,ZhouCao2025}, and BCH and cyclic
codes~\cite{Galindo2026BCH,Rajpurohit2026Cyclic}, with related algebraic
tools provided by NSC quasi-unitary matrices~\cite{CaoZhou2026NSC}.
Impure constructions demonstrate the effect of purity on achievable
parameters~\cite{Galindo2026Impure}. Extensions include
availability~\cite{Bu2025Intersecting,Stylianou2026Intersecting},
hierarchical~\cite{Guruswami2026Hierarchical} and asymmetric
locality~\cite{LiJinXing2026Asymmetric}, and entanglement-assisted
recovery~\cite{Li2026EA,KumarBandi2026EA,Galindo2026EA,MatthewsShapiro2026,KumarBandi2026AsymmetricEA}.
Information locality and measurement-efficient recovery have also
been studied~\cite{Matsumoto2026Measurements,Matsumoto2026Information,Matsumoto2026Simultaneous}.

Existing bounds differ in their structural assumptions.
For single-erasure recovery, Golowich and Guruswami established a
general quantum-subspace bound and a stronger bound for disjoint
recovery blocks~\cite{Golowich2023LRC}; Luo et al. obtained
Singleton-like and Cadambe--Mazumdar-type bounds for CSS
constructions~\cite{Luo2023CSS}.
For multiple erasures, Galindo et al. derived a bound for Euclidean
or Hermitian dual-containing classical
constructions~\cite{Galindo2026Locality}. It involves the classical
distance, which can be replaced by the actual quantum distance for
pure codes, but generally not for impure
codes~\cite{Galindo2026Impure}.
Guruswami, Kshirsagar, and Trivedi established a bound without purity
for equal-sized disjoint recovery blocks, together with bounds for
hierarchical constructions~\cite{Guruswami2026Hierarchical}.
Stylianou and Boche obtained Singleton-like and linear programming
bounds for pure quantum subspaces with disjoint recovery
blocks~\cite{StylianouBoche2026}.
Pure Hermitian constructions also admit alphabet-dependent
bounds~\cite{Li2025Pure,KumarBandi2026}, while bounds for the extended
models above depend on their additional recovery or resource
assumptions. These multiple-erasure results therefore retain
restrictions on the construction, purity, or block partition.
The general single-erasure bound permits impure codes and arbitrary
overlaps, but does not exploit the full local multiple-erasure
capability.

We establish a Singleton-like bound for general $(r,\delta)$-qLRCs,
without assuming purity, a stabilizer structure, or disjoint recovery
blocks. The proof combines a two-stage selection lemma with a coverage
budget and an entropy argument. At $\delta=2$, the bound reduces to
the general Golowich--Guruswami bound~\cite{Golowich2023LRC};
integer optimization yields an explicit correction of up to
$\delta-2$. Under purity, local decoupling gives a stronger bound,
extending the pure-code expression of
Galindo et al.~\cite{Galindo2026Locality} to arbitrary pure quantum
subspaces and removing the partition assumption from related
Stylianou--Boche bounds~\cite{StylianouBoche2026}.
We also derive equality conditions that constrain block
intersections and force local MDS structure; when $r=\delta$,
blocks meeting the residual set in the correctable partition are
isolated. For $\mathbb F_q$-linear stabilizer codes with $r=\delta$,
we prove a decomposition into disjoint MDS inner codes and an outer
stabilizer code after removing fixed pure-state coordinate factors,
without assuming equality in the global bound. This gives distance
divisibility and an attainable dimension bound.
Finally, three families illustrate attainment:
pure codes with unavoidable overlaps and $d=\delta$ exhibit
parameter ranges of strict improvement over the general bound and
of simultaneous attainment of both bounds;
impure codes with disjoint blocks attain the general bound;
and an existing pure overlapping family with
$d>\delta$~\cite{zhou2025optimalquantumrdeltalocallyrepairable,ZhouCao2025} attains both the general and pure-code
bounds.

The remainder of the paper is organized as follows. Section~\ref{sec:preliminaries} introduces quantum local recovery, erasure correctability, and the entropy properties used throughout. Section~\ref{ch:bound} establishes the multiple-erasure selection lemma and proves the general bound, its integer optimization, and the stronger pure-code bound. Section~\ref{sec:equality-structure} analyzes equality conditions and characterizes the local structure and linear stabilizer decomposition when $r=\delta$. Section~\ref{sec:comparison} compares our results with existing parameter bounds under their respective assumptions. Section~\ref{sec:constructions} presents and analyzes attaining families with different purity and overlap properties. The full proof of the minimal-locality structure theorem and a multiple-block construction with a shared core are given in the appendices.

\section{Preliminaries}\label{sec:preliminaries}
\subsection{Notation, Quantum Channels, and Entropy}
Let $q\ge2$ be an integer; for finite-field constructions, assume that $q$ is a prime power. Write $[n]=\{1,\ldots,n\}$ and $\mathcal H=(\mathbb C^q)^{\otimes n}$. For $A\subseteq[n]$, let $\mathcal H_A=\bigotimes_{j\in A}\mathbb C^q$ and $A^c=[n]\setminus A$, and denote the identity operator on $\mathcal H_A$ by $I_A$. Let $\mathcal B(\mathcal H_A)$ denote the space of linear operators on $\mathcal H_A$. A density operator is a positive semidefinite operator of trace one. We write $\mathcal D(Q)$ for the density operators supported on the code space $Q$, namely, the states satisfying $\operatorname{supp}\rho\subseteq Q$. More explicitly, if a density operator has spectral decomposition $\rho=\sum_{j:\lambda_j>0}\lambda_j|\psi_j\rangle\langle\psi_j|$, then its support is $\operatorname{supp}\rho =\operatorname{span}\{|\psi_j\rangle:\lambda_j>0\}$. The reduced state of a joint state $\omega_{AB}$ on $A$ is $\omega_A=\operatorname{Tr}_B\omega_{AB}$, where the partial trace satisfies
\begin{equation}\label{eq:prelim-partial-trace}
 \operatorname{Tr}_A\!\left(X_A\operatorname{Tr}_B Y_{AB}\right)
 =\operatorname{Tr}_{AB}\!\left((X_A\otimes I_B)Y_{AB}\right).
\end{equation}
Here $X_A\in\mathcal B(\mathcal H_A)$ and $Y_{AB}\in\mathcal B(\mathcal H_A\otimes\mathcal H_B)$ are arbitrary linear operators on $A$ and $AB$, respectively, and $I_B$ is the identity operator on $B$.
All Hilbert spaces in this paper are finite dimensional. We use
$\mathcal H_\varnothing=\mathbb C$ and $I_\varnothing=1$.
Tensor factors indexed by coordinate sets are always placed in their
original coordinate positions, with the canonical permutation of factors
understood when necessary. For a subspace $Q\subseteq\mathcal H$ with
orthogonal projector $P$, we identify $\mathcal B(Q)$ with
$\{X\in\mathcal B(\mathcal H):PXP=X\}$ by extension by zero on $Q^\perp$.
We write $\mathcal D(\mathcal H_X)$ for the density operators on
$\mathcal H_X$, and $\operatorname{id}_X$ for the identity map on
$\mathcal B(\mathcal H_X)$; this map is distinct from the operator $I_X$.

A \textit{quantum channel} is a complex-linear, completely positive,
trace-preserving (CPTP) map between the stated operator
spaces~\cite[Section~2.2]{Watrous2018TQI}. A channel may have different
input and output registers. Tensoring it with an identity channel
means that the corresponding registers are left untouched.

\begin{definition}[Purification{\cite[Chapter~2]{Watrous2018TQI}}]
\label{chunhua}
Let $\omega_A$ be a density operator on $A$. If an auxiliary system $R$ and a unit vector $|\Psi\rangle_{AR}$ on $AR$ satisfy
$
\operatorname{Tr}_R
\bigl(|\Psi\rangle_{AR}\langle\Psi|\bigr)
=\omega_A,
$
then $|\Psi\rangle_{AR}$ is called a \textit{purification} of $\omega_A$, and $R$ is called the purifying system. Every density operator admits a purification.
\end{definition}
\begin{lemma}[Product decomposition with a pure marginal]
\label{lem:prelim-pure-marginal}
Let $\omega_{XY}$ be a density operator on a finite-dimensional bipartite system $XY$, with marginals $\omega_X=\operatorname{Tr}_Y\omega_{XY}, \; \omega_Y=\operatorname{Tr}_X\omega_{XY}$. If $\omega_X$ is pure, then
$
\omega_{XY}=\omega_X\otimes\omega_Y.
$
\end{lemma}
\begin{proof}
Purify $\omega_{XY}$ to $|\Psi\rangle_{XYR}$ and take a Schmidt decomposition across $X|(YR)$. Since $\omega_X$ has rank one, this decomposition has a single term, $|\Psi\rangle=|\phi\rangle_X|\eta\rangle_{YR}$. Taking the partial trace over $R$ proves the claim; see also~\cite[Chapter~2]{Watrous2018TQI}.
\end{proof}

\begin{definition}[Von Neumann entropy and quantum mutual information
{\cite[Section~5.2]{Watrous2018TQI}}]\label{def:prelim-entropy}
Let $X$ and $Y$ be finite-dimensional quantum systems with joint state $\omega_{XY}$ and reduced states $\omega_X=\operatorname{Tr}_Y\omega_{XY}$ and $\omega_Y=\operatorname{Tr}_X\omega_{XY}$. If the eigenvalues of $\omega_X$ are $\lambda_1,\ldots,\lambda_m$, its \textit{von Neumann entropy} with logarithms to base $q$ is
\[
 S(X)=S(\omega_X)
 =-\operatorname{Tr}_X(\omega_X\log_q\omega_X)
 =-\sum_{j=1}^{m}\lambda_j\log_q\lambda_j,
\]
with the convention $0\log_q0=0$. The \textit{quantum mutual information} of $\omega_{XY}$ is
\begin{equation}\label{eq:prelim-mutual-information}
 I(X:Y)=S(X)+S(Y)-S(XY).
\end{equation}
\end{definition}

\begin{lemma}[Basic entropy properties
{\cite[Section~5.2]{Watrous2018TQI}}]\label{lem:prelim-entropy}
Finite-dimensional quantum systems satisfy the following properties.
\begin{enumerate}[label=\textup{(\roman*)}]
 \item For every state $\omega_X$, $0\le S(X)\le\log_q\dim\mathcal H_X$. Equality holds on the left if and only if $\omega_X$ is pure, and on the right if and only if $\omega_X=I_X/\dim\mathcal H_X$ is maximally mixed.
 \item For every joint state $\omega_{XY}$, $S(XY)\le S(X)+S(Y)$. Equality holds if and only if $\omega_{XY}=\omega_X\otimes\omega_Y$. Consequently, $I(X:Y)\ge0$, with equality if and only if the joint state is a product state.
 \item If $|\Psi\rangle_{XY}$ is pure, its two reduced states have the same nonzero eigenvalues, and hence $S(X)=S(Y)$.
\end{enumerate}
\end{lemma}
The following data processing inequality states that a quantum channel acting on one subsystem cannot increase its mutual information with the remaining system.
\begin{lemma}[Data processing inequality for mutual information
{\cite[Theorem~5.35]{Watrous2018TQI}}]\label{lem:prelim-dpi}
Let $\mathcal N:\mathcal B(\mathcal H_Y)\to\mathcal B(\mathcal H_Z)$ be a quantum channel, that is, a completely positive trace-preserving linear map. If $\sigma_{XZ}=(\operatorname{id}_X\otimes\mathcal N)(\omega_{XY})$, then
\begin{equation}\label{eq:prelim-dpi}
 I(X:Z)_\sigma\le I(X:Y)_\omega.
\end{equation}
\end{lemma}

\subsection{Erasures and Local Recovery}
A quantum code $Q$ is a $K$-dimensional subspace of
$\mathcal H$. Denote its orthogonal projector by $P$ and
let $k=\log_q K$. We denote the code parameters by
$((n,K,d))_q$, where $K=\dim Q$ and $d$ is the minimum
distance, and state dimension bounds in terms of $k$.
A code is called nontrivial when $K>1$.
For an $\mathbb F_q$-linear stabilizer code, we also use
$\llbracket n,k,d\rrbracket_q$, with $K=q^k$ and integer $k$.
A quantum code of distance $d$ corrects any set of at most
$d-1$ erasures at known locations~\cite{KnillLaflamme1997}.

On each coordinate, fix an orthogonal unitary error basis
$\{U_a\}_{a=0}^{q^2-1}$ with $U_0=I$.
This is a basis of $\mathcal B(\mathbb C^q)$ consisting
of unitary operators that are mutually orthogonal under
the Hilbert--Schmidt inner product:
$
 U_a^\dagger U_a=I,
$
 $\operatorname{Tr}(U_a^\dagger U_b)=q\,\delta_{ab}.
$
In particular, $\operatorname{Tr}(U_a)=0$ for $a\ne0$.
For qubits ($q=2$), the Pauli operators
$\{I,X,Y,Z\}$ form such a basis, where
\[
 X=\begin{pmatrix}0&1\\1&0\end{pmatrix},
 \qquad
 Y=\begin{pmatrix}0&-i\\i&0\end{pmatrix},
 \qquad
 Z=\begin{pmatrix}1&0\\0&-1\end{pmatrix}.
\]
Their tensor products form an operator basis on the full
system. For a tensor-product basis error
$E=U_{a_1}\otimes\cdots\otimes U_{a_n}$, define its weight by
$
 \operatorname{wt}(E)
 =\bigl|\{j\in[n]:a_j\ne0\}\bigr|,
$
the number of its nonidentity factors.
For example, $X\otimes I\otimes Z$ has weight two.

The code $Q$ is pure if $PEP=0$ for every tensor-product
basis error $E$ with $0<\operatorname{wt}(E)<d$;
otherwise, it is impure~\cite{HuberGrassl2020MDS}.
Equivalently, for every such $E$ and all
$|\phi\rangle,|\psi\rangle\in Q$,
$
 \langle\phi|E|\psi\rangle
 =\langle\phi|PEP|\psi\rangle=0.
$
Thus, purity requires every nonidentity basis error of
weight less than $d$ to map the code space into a
subspace orthogonal to it.

\begin{definition}[Correctable erasure set and local recovery channel]
\label{def:prelim-erasure}
Following~\cite[full version, Definition~25]{Golowich2023LRC},
erasure of a known set $E\subseteq[n]$ is represented by
the partial trace
$\operatorname{Tr}_E:\mathcal B(\mathcal H)\to
\mathcal B(\mathcal H_{E^c})$.
The set $E$ is \textit{correctable for $Q$} if there exists
a quantum channel
$
 \operatorname{Rec}_E:
 \mathcal B(\mathcal H_{E^c})
 \longrightarrow\mathcal B(\mathcal H)
$
such that
\begin{equation}\label{eq:prelim-exact-recovery}
 \operatorname{Rec}_E(\operatorname{Tr}_E\tau)=\tau,
 \qquad \forall\,\tau\in\mathcal D(Q).
\end{equation}

For $E\subseteq\Gamma\subseteq[n]$, a quantum channel
$
 \operatorname{Rec}_{\Gamma,E}:
 \mathcal B(\mathcal H_{\Gamma\setminus E})
 \longrightarrow\mathcal B(\mathcal H_\Gamma)
$
is called a \textit{local recovery channel within $\Gamma$}
for erasures on $E$ if
\[
 (\operatorname{Rec}_{\Gamma,E}
 \otimes\operatorname{id}_{\Gamma^c})
 (\operatorname{Tr}_E\tau)=\tau,
 \qquad \forall\,\tau\in\mathcal D(Q).
\]
Its input registers are $\Gamma\setminus E$, its output
registers are $\Gamma$, and the registers $\Gamma^c$
are left untouched.
\end{definition}

\begin{definition}[$(r,\delta)$-qLRC
{\cite{Galindo2026Locality}}]
\label{def:prelim-rdelta}
Let $r\ge1$ and $\delta\ge2$ be integers.
A quantum code $Q\subseteq(\mathbb C^q)^{\otimes n}$
is a \textit{quantum $(r,\delta)$-locally recoverable code}
if, for every $i\in[n]$, there exists a designated recovery
block $\Gamma_i\subseteq[n]$ containing $i$, with
$|\Gamma_i|\le r+\delta-1$, such that every
$E\subseteq\Gamma_i$ with $|E|\le\delta-1$ admits a local
recovery channel $\operatorname{Rec}_{\Gamma_i,E}$
within $\Gamma_i$ satisfying
\begin{equation}\label{eq:prelim-local-recovery}
 (\operatorname{Rec}_{\Gamma_i,E}
 \otimes\operatorname{id}_{\Gamma_i^c})
 (\operatorname{Tr}_E\tau)=\tau,
 \qquad \forall\,\tau\in\mathcal D(Q).
\end{equation}
\end{definition}
Each recovery channel is fixed independently of the unknown code state.
The condition for all density operators is equivalent, by linearity,
to the condition for all pure code states. Moreover, density operators
supported on $Q$ span $\mathcal B(Q)$, so these identities also preserve
all correlations with an arbitrary reference system.
When $\delta=2$, every coordinate of $\Gamma_i$ can be recovered from
the other registers in $\Gamma_i$. In particular, $Q$ has single-erasure
locality with block size at most $r+1$ in the convention
of~\cite{Golowich2023LRC}, where locality counts the whole block.
The code has \textit{disjoint recovery blocks} if any two designated
blocks are identical or disjoint~\cite{StylianouBoche2026}.

\subsection{Decoupling of Correctable Erasure Sets from the Reference System}
The following lemma applies the information-theoretic
characterization of exact quantum error correction
to erasures~\cite{SchumacherNielsen1996}.

\begin{lemma}[Correctability implies decoupling]
\label{lem:prelim-decoupling}
Let $A\subseteq[n]$ be a correctable erasure set for $Q$.
For any $\sigma\in\mathcal D(Q)$, let
$|\Psi\rangle_{RAA^c}$ be a purification of $\sigma$
with reference system $R$, and write
$\omega=|\Psi\rangle\langle\Psi|$.
Subscripts on $\omega$ denote its reduced states;
in particular,
$\omega_{RA}=\operatorname{Tr}_{A^c}\omega$.
Then
\[
 \omega_{RA}=\omega_R\otimes\omega_A,
 \qquad I(R:A)_\omega=0.
\]
\end{lemma}

\begin{proof}
Let $\operatorname{Rec}_A:\mathcal B(\mathcal H_{A^c})\to
\mathcal B(\mathcal H)$ be a recovery channel for $A$.
The identity $\operatorname{Rec}_A\circ\operatorname{Tr}_A
=\operatorname{id}$ holds on $\mathcal B(Q)$ by linearity.
Expanding $\omega$ in matrix units on $Q$ therefore gives
\[
 (\operatorname{id}_R\otimes\operatorname{Rec}_A)(\omega_{RA^c})
 =\omega.
\]
By data processing and purity of $\omega$,
\[
 I(R:A^c)_\omega
 \ge I(R:AA^c)_\omega=2S(R)_\omega.
\]
Purity also gives
\[
 I(R:A)_\omega+I(R:A^c)_\omega=2S(R)_\omega.
\]
Nonnegativity thus implies $I(R:A)_\omega=0$,
equivalently $\omega_{RA}=\omega_R\otimes\omega_A$.
\end{proof}

For the subsequent entropy arguments, let $\rho=P/K$
be the maximally mixed state on $Q$.
Choose an orthonormal basis $\{|q_j\rangle\}_{j=1}^K$
of $Q$ and a $K$-dimensional reference system $R_0$,
and fix the purification
\begin{equation}\label{eq:prelim-code-purification}
 |\Omega\rangle_{R_0[n]}
 =\frac1{\sqrt K}\sum_{j=1}^K
 |j\rangle_{R_0}\otimes|q_j\rangle,
 \qquad
 \rho_{R_0}=\frac{I_{R_0}}K,
 \qquad S(R_0)=\log_qK=k.
\end{equation}
Here, subscripts on $\rho$ denote the corresponding
reduced states of $|\Omega\rangle\langle\Omega|$.
By Lemma~\ref{lem:prelim-decoupling}, every correctable
erasure set $A$ satisfies
\begin{equation}\label{eq:prelim-decoupling}
 I(R_0:A)=0,
 \qquad
 \rho_{R_0A}=\rho_{R_0}\otimes\rho_A.
\end{equation}
\begin{lemma}[Entropy bound from a correctable erasure set]
\label{cor:prelim-correctable-entropy}
Let $A\subseteq[n]$ be a correctable erasure set for $Q$,
and let $\rho=P/K$.
Then
\begin{equation}\label{eq:prelim-correctable-entropy}
 k\le n-|A|-S(A).
\end{equation}
In particular, if $\rho_A=I_A/q^{|A|}$, then $k\le n-2|A|$.
\end{lemma}
\begin{proof}
Consider the purification
$|\Omega\rangle_{R_0[n]}$ of $\rho=P/K$
in~\eqref{eq:prelim-code-purification}, for which
$S(R_0)=k$.
Since $A$ is correctable,
\eqref{eq:prelim-decoupling} gives
$\rho_{R_0A}=\rho_{R_0}\otimes\rho_A$, and hence
$
 S(R_0A)=S(R_0)+S(A)=k+S(A).
$
The global state on $R_0AA^c$ is pure, so
\[
 k+S(A)=S(R_0A)=S(A^c)
 \le\log_q\dim\mathcal H_{A^c}=n-|A|.
\]
Rearranging proves~\eqref{eq:prelim-correctable-entropy}.
If $\rho_A=I_A/q^{|A|}$, then $S(A)=|A|$,
giving $k\le n-2|A|$.
\end{proof}

\begin{lemma}[Two disjoint correctable erasure sets
{\cite{Bu2025Intersecting}}]
\label{lem:prelim-bu}
Let $Q\subseteq(\mathbb C^q)^{\otimes n}$ be a quantum code
with $K=\dim Q$ and $k=\log_q K$.
If $A_1,A_2\subseteq[n]$ are disjoint correctable erasure sets
for $Q$, then
\begin{equation}\label{eq:prelim-bu}
 k\le n-|A_1|-|A_2|.
\end{equation}
\end{lemma}

\section{General and Pure-Code Bounds with Arbitrarily Overlapping Recovery Blocks}\label{ch:bound}
\subsection{Local Support Codes and Recovery-Block Size}
We use the local recovery model of Definition~\ref{def:prelim-rdelta}, with erasure channels as in Definition~\ref{def:prelim-erasure}. Write $t=\delta-1$, $b=r+t$, and $D=d-1$, and assume $K=q^k>1$.

\begin{figure}[htbp]
 \centering
 \includegraphics[width=0.78\linewidth]{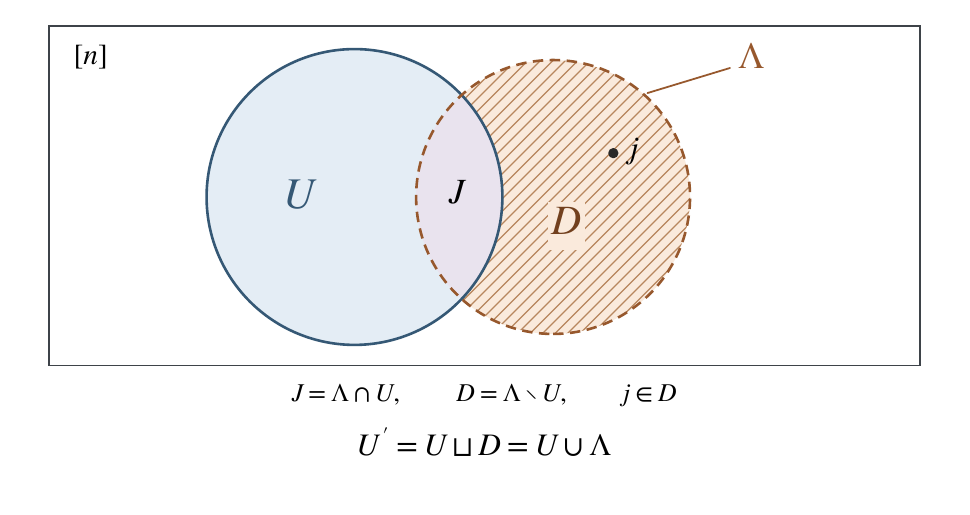}
 \caption{An induction step in the proof of
 Lemma~\ref{lem:theory-blocksize}.
 The rectangle denotes $[n]$.
 The current set $U$ and the recovery block $\Lambda$
 intersect in $J$; the hatched region is
 $D=\Lambda\setminus U$ and contains the chosen coordinate $j$.
 The current factorization gives
 $\rho_\Lambda=\rho_J\otimes\rho_D$.
 Since $\rho_\Lambda$ is pure, $\rho_D$ is pure,
 allowing $U$ to be enlarged to $U'=U\sqcup D$.
 Region areas do not represent coordinate counts.}
 \label{fig:blocksize-induction}
\end{figure}

To study local multiple-erasure correctability, we first translate recovery on a block into an error-correction condition for its local support subspace. Stylianou and Boche~\cite[Definition~3 and Theorem~1]{StylianouBoche2026} introduced local support codes and characterized recovery within a block through local Knill--Laflamme conditions. This local criterion does not depend on disjointness between recovery blocks, and its proof does not require purity. It therefore applies to the general local recovery model considered here. For completeness, we restate the definition and equivalence and give a proof for erasure channels.
\begin{definition}[Local support code]\label{def:local-support}
Let $Q\subseteq\mathcal H$ be a $K$-dimensional quantum code with orthogonal projector $P$, and let $\rho=P/K$ be the maximally mixed state on the code space. For $\Gamma\subseteq[n]$, write $\rho_\Gamma=\operatorname{Tr}_{\Gamma^c}\rho$ for its reduced state on $\Gamma$. Define
$
Q_\Gamma
=\operatorname{supp}\rho_\Gamma
\subseteq\mathcal H_\Gamma,
$
which is called the \textit{local support code} of $Q$ on $\Gamma$.  
\end{definition}
\begin{lemma}[Erasure correctability of the local support code]
\label{lem:theory-support}
Let $Q_\Gamma$ be the local support code of $Q$ on $\Gamma$, and let
$E\subseteq\Gamma$. A recovery channel acting only on $\Gamma$ corrects
erasures on $E$ for every state in $Q$ if and only if $Q_\Gamma$
corrects erasures on $E$.
\end{lemma}
\begin{proof}
Write $B=\Gamma^c$, $\rho=P/K$, and
$\rho_\Gamma=\operatorname{Tr}_B\rho$.
Suppose first that a block recovery channel
$\operatorname{Rec}_{\Gamma,E}$ satisfies
\eqref{eq:prelim-local-recovery}, and put
\[
 T=\operatorname{Rec}_{\Gamma,E}\circ\operatorname{Tr}_E,
 \qquad F=T\otimes\operatorname{id}_B.
\]
Here $\operatorname{Tr}_E$ maps $\mathcal B(\mathcal H_\Gamma)$ to
$\mathcal B(\mathcal H_{\Gamma\setminus E})$, so $T$ is a channel
on $\mathcal B(\mathcal H_\Gamma)$.
The channel $F$ fixes $\mathcal B(Q)$ by linearity. Expanding a
purification $|\Omega\rangle_{\Gamma BR}$ of $\rho$ in code matrix
units gives
\begin{equation}\label{eq:local-support-fix-purification}
 (T\otimes\operatorname{id}_{BR})(|\Omega\rangle\langle\Omega|)
 =|\Omega\rangle\langle\Omega|.
\end{equation}
Take a Schmidt decomposition across $\Gamma|(BR)$,
\[
 |\Omega\rangle=\sum_{a=1}^{d_\Gamma}\sqrt{\lambda_a}\,
 |a\rangle_\Gamma\otimes|e_a\rangle_{BR},
 \qquad d_\Gamma=\operatorname{rank}\rho_\Gamma,\quad\lambda_a>0.
\]
The vectors $|a\rangle$ form an orthonormal basis of $Q_\Gamma$.
Taking the $BR$ matrix element of
\eqref{eq:local-support-fix-purification} between
$\langle e_a|$ and $|e_b\rangle$ gives
\[
 \sqrt{\lambda_a\lambda_b}\,T(|a\rangle\langle b|)
 =\sqrt{\lambda_a\lambda_b}\,|a\rangle\langle b|.
\]
All Schmidt coefficients are positive, so $T$ fixes every matrix
unit of $\mathcal B(Q_\Gamma)$. In particular, the same CPTP map
$\operatorname{Rec}_{\Gamma,E}$, with input registers
$\Gamma\setminus E$ and output registers $\Gamma$, recovers every state in
$\mathcal D(Q_\Gamma)$ after erasure on $E$.

Conversely, suppose that $Q_\Gamma$ corrects erasures on $E$, and
choose a recovery channel
$\operatorname{Rec}_{\Gamma,E}:\mathcal B(\mathcal H_{\Gamma\setminus E})
\to\mathcal B(\mathcal H_\Gamma)$. Set
$T=\operatorname{Rec}_{\Gamma,E}\circ\operatorname{Tr}_E$.
We have the support inclusion
\begin{equation}\label{eq:local-support-code-inclusion}
 Q\subseteq Q_\Gamma\otimes\mathcal H_B.
\end{equation}
Indeed, let $\Pi_\Gamma$ project onto $Q_\Gamma$ and set
$A=(I_\Gamma-\Pi_\Gamma)\otimes I_B$. Then
$\operatorname{Tr}(A\rho)=
\operatorname{Tr}((I_\Gamma-\Pi_\Gamma)\rho_\Gamma)=0$.
Since $A$ is a projector and $\rho\ge0$, this implies
$A\rho^{1/2}=0$, proving~\eqref{eq:local-support-code-inclusion}.
Since $T$ fixes every state in $\mathcal D(Q_\Gamma)$, it fixes
$\mathcal B(Q_\Gamma)$ by linearity. Every $\tau\in\mathcal D(Q)$ is
supported on $Q_\Gamma\otimes\mathcal H_B$, so expanding $\tau$ in
matrix units on $Q_\Gamma$ yields
\[
 (T\otimes\operatorname{id}_B)(\tau)=\tau.
\]
Thus $\operatorname{Rec}_{\Gamma,E}$, extended by the identity on $B$,
corrects erasures on $E$ for every state in $Q$.
\end{proof}
The preceding lemma reduces recovery within a block to
erasure correction on its local support code.
Applying Lemma~\ref{lem:prelim-bu} to a partition of the block
into two correctable sets constrains its local support
dimension.
The next lemma uses this constraint to show that every
nontrivial code must satisfy $b\ge2t+1$, even when its
recovery blocks overlap.
\begin{lemma}[Block-size constraint for nontrivial codes]\label{lem:theory-blocksize}
If a quantum $(r,\delta)$-qLRC with $K>1$ exists, then $b\ge2t+1$.
\end{lemma}
\begin{proof}
Suppose, to the contrary, that $b\le 2t$. Let $\rho=P/K$, where $P$ is the orthogonal projector onto $Q$, and write $\rho_S=\operatorname{Tr}_{S^c}\rho$.

We first show that the marginal on each designated recovery
block $\Gamma$ is pure.
By Lemma~\ref{lem:theory-support},
$Q_\Gamma=\operatorname{supp}\rho_\Gamma$ corrects any set
of at most $t$ erasures in $\Gamma$.
Since $|\Gamma|\le b\le2t$, partition
$\Gamma=A\sqcup B$ with $|A|,|B|\le t$.
Both $A$ and $B$ are therefore correctable erasure sets
for $Q_\Gamma$.
Applying Lemma~\ref{lem:prelim-bu} to this local code gives
$
 \log_q\dim Q_\Gamma
 \le|\Gamma|-|A|-|B|=0.
$
Thus $\dim Q_\Gamma=1$, and $\rho_\Gamma$ is pure.

Choose a designated recovery block $\Gamma_1$. Starting with $U=\Gamma_1$, Lemma~\ref{lem:prelim-pure-marginal} gives $\rho=|\phi_U\rangle\langle\phi_U|\otimes\rho_{U^c}$. We enlarge $U$ inductively while preserving this factorization. If $U\ne[n]$, choose $j\in U^c$ and a designated recovery block $\Lambda$ containing $j$. Set $J=\Lambda\cap U,\; D=\Lambda\setminus U$. Then $\Lambda=J\sqcup D$, and $j\in D$ ensures $D\ne\varnothing$. Taking the marginal on $\Lambda$ in the current factorization gives
\[
\begin{aligned}
\rho_\Lambda
&=
\left(\operatorname{Tr}_{U\setminus J}
|\phi_U\rangle\langle\phi_U|\right)
\otimes
\left(\operatorname{Tr}_{U^c\setminus D}\rho_{U^c}\right)
=\rho_J\otimes\rho_D,
\end{aligned}
\]
with the convention $\rho_\varnothing=1$. Since $\rho_\Lambda$ is pure, $1=\operatorname{rank}\rho_\Lambda =\operatorname{rank}\rho_J\, \operatorname{rank}\rho_D$, so $\rho_D$ is also pure; write $\rho_D=|\phi_D\rangle\langle\phi_D|$.
Applying Lemma~\ref{lem:prelim-pure-marginal} again, now to the pure marginal of $\rho_{U^c}$ on $D$, gives $\rho_{U^c} =|\phi_D\rangle\langle\phi_D| \otimes\rho_{U^c\setminus D}$. Let $U'=U\cup D$. Substitution into the original factorization yields
\[
\begin{aligned}
\rho
&=|\phi_U\rangle\langle\phi_U|
  \otimes|\phi_D\rangle\langle\phi_D|
  \otimes\rho_{(U')^c}
=|\phi_{U'}\rangle\langle\phi_{U'}|
  \otimes\rho_{(U')^c},
\end{aligned}
\qquad
|\phi_{U'}\rangle=|\phi_U\rangle\otimes|\phi_D\rangle.
\]
Because $\varnothing\ne D\subseteq U^c$, we have $U\subsetneq U'$, so each step adds at least one new coordinate.
After finitely many steps, $U=[n]$, making $\rho$ pure. This contradicts $\operatorname{rank}\rho=\operatorname{rank}(P/K)=K>1$. Therefore $b>2t$, and integrality gives $b\ge2t+1$.
\end{proof}
In particular, every nontrivial code in this model satisfies $r\ge\delta$.
Figure~\ref{fig:blocksize-induction} illustrates this
enlargement: the recovery block $\Lambda$ meets $U$ in $J$
and adds the nonempty set $D=\Lambda\setminus U$.

\subsection{Selecting Multiple-Erasure Sets}
The single-erasure bound of Golowich and Guruswami constructs correctable erasure sets by selecting recovery blocks sequentially~\cite{Golowich2023LRC}. Incremental block-selection arguments for classical locality bounds account for overlaps through newly added coordinates and constraints~\cite{Gopalan2012Locality,prakash2012optimallinearcodeslocalerrorcorrection}. Starting from local multiple-erasure recovery channels, the following lemma selects coordinates that can be recovered sequentially within a prescribed coverage budget, while ensuring that the recovery blocks used avoid additional erasures outside the covered region.

\begin{figure}[htbp]
  \centering
  \includegraphics[width=\linewidth]{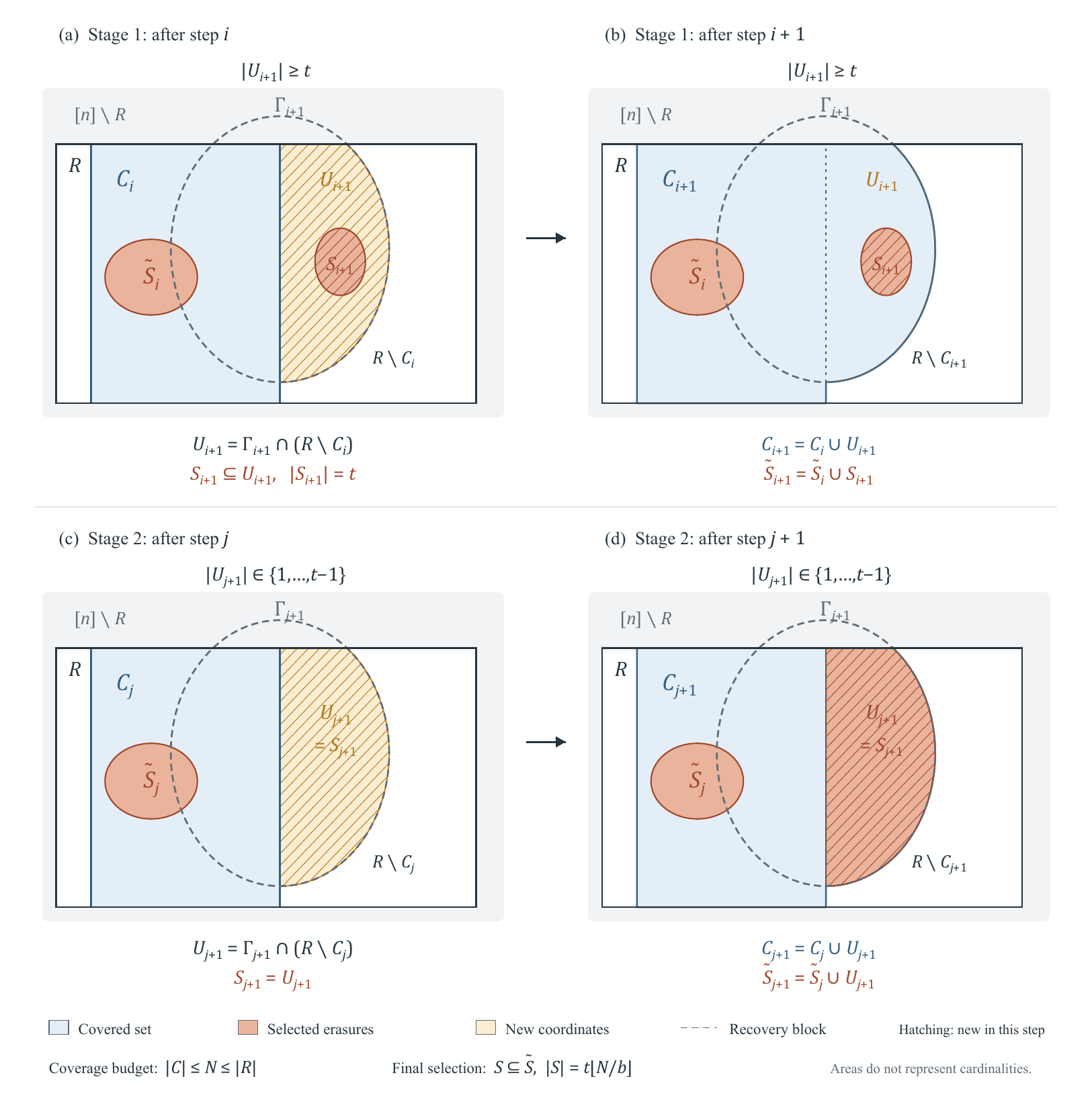}
  \caption{Updates in the first stage (top) and second stage (bottom). Blue denotes covered coordinates, red denotes selected erasures, and dashed ellipses denote recovery blocks. At each step, all new coordinates of the block in $R$ are covered. The first stage selects $t$ of them as erasures, whereas the second stage selects all of them. Region areas do not represent coordinate counts.}
  \label{fig:erasure-selection}
\end{figure}

The additional step addresses the following obstacle: after previously selected blocks have been covered, every remaining block may contain fewer than $t$ uncovered coordinates. Simply replacing the selection of one coordinate per step in a single-erasure argument by the selection of $t$ coordinates can therefore cause the process to stop prematurely. The second stage below uses these smaller increments and shows that the total coverage remains within the original budget.
\begin{lemma}[Multiple-erasure set selection]\label{lem:theory-selection}
Suppose that every coordinate has a designated recovery block of size at most $b$, any set of at most $t$ erasures in a block can be recovered within that block, and $b\ge2t+1$. For every $R\subseteq[n]$ and integer $0\le N\le|R|$, there exist sets $S\subseteq C\subseteq R$ such that
\[
 |S|=t\left\lfloor N/b\right\rfloor,\qquad |C|\le N,
\]
with the following property: all blocks needed to recover $S$ avoid $R\setminus C$, and local recovery can be performed in a finite sequence. Even if an arbitrary set $V\subseteq R\setminus C$ is erased simultaneously, $S$ can be recovered first, leaving only $V$ erased for subsequent recovery.
\end{lemma}
\begin{proof}
Write $M=\lfloor N/b\rfloor$. If $M=0$, take $S=C=\varnothing$. Assume henceforth that $M\ge1$. We first construct $\widetilde S\subseteq C\subseteq R$ with $|\widetilde S|\ge tM$ and $|C|\le N$, and then reduce $\widetilde S$ to the required size.

Initialize $C=\widetilde S=\varnothing$. For a designated recovery block $\Gamma$, let $U=\Gamma\cap(R\setminus C)$ be its uncovered coordinates. The first stage consists of at most $M$ steps. Whenever a block satisfies $|U|\ge t$, select $S_i\subseteq U$ with $|S_i|=t$, and update
\[
\widetilde S\longleftarrow\widetilde S\cup S_i,
\qquad
C\longleftarrow C\cup U.
\]
Each step covers at most $b$ new coordinates. If $M$ steps are completed, then $|\widetilde S|=tM,\; |C|\le bM\le N$. Otherwise, the first stage stops after $a<M$ steps. Denote the covered set at this point by $C_a$. Then $|\widetilde S|=at$, $|C_a|\le ab$, and every designated block satisfies $|\Gamma\cap(R\setminus C_a)|<t$. Since $C$ only grows, this property persists. If $t=1$, then $|C_a|\le ab<bM\le N\le|R|$, so an uncovered coordinate remains, and its designated block contains at least one new coordinate. This contradicts termination of the first stage. Thus, only $t\ge2$ needs to be considered in this case.

In the second stage, while $|\widetilde S|<tM$, choose $j\in R\setminus C$ and a designated block $\Gamma(j)$ containing $j$. Then $U=\Gamma(j)\cap(R\setminus C), \; 1\le|U|\le t-1$. Set $S_i=U$ and add $U$ to both $\widetilde S$ and $C$. Throughout the second stage, $|C|=|C_a|+|\widetilde S|-at \le ab+|\widetilde S|-at$. If $C=R$ before the target is reached, then
\[
\begin{aligned}
|R|
\le ab+|\widetilde S|-at
<ab+t(M-a)
&=bM-(b-t)(M-a)
\le bM\le N\le|R|,
\end{aligned}
\]
a contradiction. Hence a new coordinate is always available before the target is reached. Each step adds at least one coordinate, so the process terminates after finitely many steps.

The last step adds at most $t-1$ coordinates. Thus, at termination, $tM\le|\widetilde S|\le tM+t-2$. Using $b-t\ge t+1$ and $M-a\ge1$, we obtain
\[
\begin{aligned}
|C|
&\le ab+t(M-a)+t-2
=bM-(b-t)(M-a)+t-2
\le bM\le N.
\end{aligned}
\]
Whether or not the first stage completes $M$ steps, we therefore have
\[
\widetilde S\subseteq C\subseteq R,
\qquad |\widetilde S|\ge tM,\qquad |C|\le N.
\]

We next verify local recoverability. List the selected blocks as $\Gamma_1,\ldots,\Gamma_L$, the corresponding erasure sets as $S_1,\ldots,S_L$, and the covered set after step $i$ as $C_i$. By construction, the $S_i$ are pairwise disjoint and
\[
\widetilde S=\bigsqcup_{i=1}^{L}S_i,
\qquad
S_i\subseteq\Gamma_i,\quad |S_i|\le t,
\qquad
\Gamma_i\cap R\subseteq C_i.
\]
For $h>i$, $S_h\subseteq R\setminus C_{h-1} \subseteq R\setminus C_i$, so $S_h\cap\Gamma_i=\varnothing$. Moreover, for any $V\subseteq R\setminus C$, the inclusion $\Gamma_i\cap R\subseteq C_i\subseteq C$ gives $V\cap\Gamma_i=\varnothing$.

For each $i$, choose a local recovery channel
$\operatorname{Rec}_{\Gamma_i,S_i}$ as in
Definition~\ref{def:prelim-rdelta}. Write
$W_i=V\cup\bigcup_{h>i}S_h$ and $B_i=\Gamma_i^c\setminus W_i$.
Since $W_i\subseteq\Gamma_i^c$, the channel
$\operatorname{Rec}_{\Gamma_i,S_i}\otimes\operatorname{id}_{B_i}$
has input registers $[n]\setminus(S_i\cup W_i)$ and output registers
$[n]\setminus W_i$. Taking the partial trace over $W_i$ in the
local recovery identity gives, for every $\tau\in\mathcal D(Q)$,
\begin{equation}\label{eq:selection-trace-recovery}
 \begin{aligned}
 &(\operatorname{Rec}_{\Gamma_i,S_i}\otimes\operatorname{id}_{B_i})
   (\operatorname{Tr}_{S_i\cup W_i}\tau)
=\operatorname{Tr}_{W_i}\!\left[
   (\operatorname{Rec}_{\Gamma_i,S_i}\otimes
     \operatorname{id}_{\Gamma_i^c})(\operatorname{Tr}_{S_i}\tau)
   \right]
 =\operatorname{Tr}_{W_i}\tau.
 \end{aligned}
\end{equation}
Here the partial trace passes through the recovery channel
because $W_i$ is outside $\Gamma_i$; the identity factor
acts only on the registers $B_i$ that remain present.
Applying the recovery channels in the selection order gives
\[
 \begin{aligned}
 &\bigl(\operatorname{Rec}_{\Gamma_\ell,S_\ell}
          \otimes\operatorname{id}_{B_\ell}\bigr)
 \circ\cdots\circ
 \bigl(\operatorname{Rec}_{\Gamma_1,S_1}
          \otimes\operatorname{id}_{B_1}\bigr)
 \bigl(\operatorname{Tr}_{\widetilde S\cup V}\tau\bigr)
=\operatorname{Tr}_V\tau,
 \qquad \forall\,\tau\in\mathcal D(Q).
 \end{aligned}
\]
Their composition is a quantum channel from
$\mathcal B(\mathcal H_{[n]\setminus(\widetilde S\cup V)})$
to $\mathcal B(\mathcal H_{[n]\setminus V})$.
The same local maps
$\operatorname{Rec}_{\Gamma_i,S_i}$ work for every
$V\subseteq R\setminus C$, with the identity factors
chosen according to the remaining registers.
Coordinates of the recovery blocks outside $R$ remain
unerased throughout and can be used in these operations.

Finally, choose any $S\subseteq\widetilde S$ with $|S|=tM$ and set
$S_i'=S_i\cap S$. Keep the original selection order and omit empty
$S_i'$. We still have $S_i'\subseteq\Gamma_i$, $|S_i'|\le t$, and
\[
 \left(V\cup\bigcup_{h>i}S_h'\right)\cap\Gamma_i=\varnothing.
\]
Choose the block recovery channels for these smaller erasure sets.
The same induction recovers $S$ from $\operatorname{Tr}_{S\cup V}\tau$
and leaves exactly $\operatorname{Tr}_V\tau$. Thus,
\[
 S\subseteq C\subseteq R,\qquad
 |S|=t\left\lfloor\frac Nb\right\rfloor,\qquad |C|\le N,
\]
and every recovery block used avoids $R\setminus C$.
\end{proof}
Fig.~\ref{fig:erasure-selection} illustrates the two types of set updates in Lemma~\ref{lem:theory-selection}.

\subsection{A Universal Dimension Bound}
The selection method above allows the selected erasures to be recovered locally before the remaining erasures are handled using the global distance. We use two rounds of selection to construct two disjoint correctable erasure sets and derive a dimension bound from their decoupling from the reference system.
\begin{theorem}[Multiple-erasure Singleton-like bound without a purity assumption]
\label{thm:theory-bound}
Let $Q$ be an $((n,K,d))_q$ quantum $(r,\delta)$-qLRC with $K=q^k>1$ and $\delta\ge2$. Then
\begin{equation}\label{eq:theory-budget-nonnegative}
n-2d+2
-(\delta-1)
\left\lfloor
\frac{n-d+1}{r+\delta-1}
\right\rfloor
\ge0,
\end{equation}
and
\begin{equation}\label{eq:theory-main-bound}
\begin{aligned}
k\le{}&
n-2d+2
-(\delta-1)
\left\lfloor
\frac{n-d+1}{r+\delta-1}
\right\rfloor
-(\delta-1)
\left\lfloor
\frac{
n-2d+2
-(\delta-1)
\left\lfloor
\frac{n-d+1}{r+\delta-1}
\right\rfloor
}{
r+\delta-1
}
\right\rfloor.
\end{aligned}
\end{equation}
In particular, if the parameters make the left-hand side of~\eqref{eq:theory-budget-nonnegative} negative, no such code with $K>1$ exists; this negative value is not to be substituted into the second floor term.
\end{theorem}
\begin{proof}
Write $t=\delta-1$, $b=r+t$, and $D=d-1$.
Lemma~\ref{lem:theory-blocksize} gives $b\ge2t+1$,
and the quantum Singleton bound gives $n\ge2D$.
Apply Lemma~\ref{lem:theory-selection} with $R=[n]$ and
$N=n-D$ to obtain $S_1\subseteq C_1$ with
$|S_1|=t\lfloor(n-D)/b\rfloor$ and $|C_1|\le n-D$.
Choose $V_1\subseteq[n]\setminus C_1$ of size $D$ and set
$A_1=S_1\sqcup V_1$. Local recovery of $S_1$, followed by
distance-based recovery of $V_1$, makes $A_1$ correctable, with
\begin{equation}\label{eq:theory-A1-size}
 |A_1|=D+t\left\lfloor\frac{n-D}{b}\right\rfloor.
\end{equation}

Set $R_2=[n]\setminus A_1$.
If $|R_2|\le D$, then $R_2$ is also correctable, and
Lemma~\ref{lem:prelim-bu} applied to $A_1,R_2$ gives $k\le0$,
a contradiction. Hence $N_2=|R_2|-D\ge0$,
which proves~\eqref{eq:theory-budget-nonnegative}.
Apply Lemma~\ref{lem:theory-selection} to $R_2$ with budget $N_2$
to obtain $S_2\subseteq C_2\subseteq R_2$, where
$|S_2|=t\lfloor N_2/b\rfloor$ and $|C_2|\le N_2$.
Choose $V_2\subseteq R_2\setminus C_2$ of size $D$.
Then $A_2=S_2\sqcup V_2$ is correctable by the same argument
and is disjoint from $A_1$.
The recovery of $A_2$ may use coordinates in $A_1$, which
remain available when $A_2$ alone is erased.
Lemma~\ref{lem:prelim-bu} now gives
\[
 k\le n-|A_1|-|A_2|
 =n-2D-t\left\lfloor\frac{n-D}{b}\right\rfloor
       -t\left\lfloor\frac{N_2}{b}\right\rfloor.
\]
Substituting $N_2=n-2D-t\lfloor(n-D)/b\rfloor$
proves~\eqref{eq:theory-main-bound}.
\end{proof}
The first-round local erasure set can be reduced, allowing the allocation of coordinates between the two rounds to be adjusted.
\begin{corollary}[Integer optimization]\label{cor:theory-optimization}
Under the assumptions of Theorem~\ref{thm:theory-bound},
\begin{equation}\label{eq:theory-opt}
k\le n-2d+2
-
\max_{\substack{
a\in\mathbb Z\\
0\le a\le
(\delta-1)\left\lfloor
\frac{n-d+1}{r+\delta-1}
\right\rfloor
}}
\left\{
a+(\delta-1)
\left\lfloor
\frac{n-2d+2-a}{r+\delta-1}
\right\rfloor
\right\}.
\end{equation}
\end{corollary}
\begin{proof}
Write $t=\delta-1$, $b=r+t$, and $D=d-1$.
For any integer $0\le a\le t\lfloor(n-D)/b\rfloor$,
shrink the first-round set $S_1$ in the preceding proof
to $S_1^{(a)}$ of size $a$.
The set $A_1^{(a)}=S_1^{(a)}\sqcup V_1$ remains correctable
and has size $D+a$.
By~\eqref{eq:theory-budget-nonnegative}, $N=n-2D-a\ge0$.
Apply Lemma~\ref{lem:theory-selection} to
$R=[n]\setminus A_1^{(a)}$ with budget $N$ and add
$D$ coordinates outside the covered set.
This yields a correctable set $A_2^{(a)}\subseteq R$ of size
$D+t\lfloor N/b\rfloor$.
Lemma~\ref{lem:prelim-bu} gives
\[
 k\le n-|A_1^{(a)}|-|A_2^{(a)}|
 =n-2D-a-t\left\lfloor\frac{n-2D-a}{b}\right\rfloor.
\]
Optimizing over $a$ proves~\eqref{eq:theory-opt}.
\end{proof}

\begin{proposition}[Exact integer correction]\label{prop:closed-optimized}
Under the assumptions of Theorem~\ref{thm:theory-bound}, let
\[
t=\delta-1,\quad b=r+t,\quad L=n-2d+2,\quad
A=t\left\lfloor\frac{n-d+1}{b}\right\rfloor,
\qquad L-A=bu+v,\quad 0\le v<b.
\]
The right-hand side of Corollary~\ref{cor:theory-optimization} is
\begin{equation}\label{eq:closed-optimized}
F_* = L-A-tu-\varepsilon,\qquad
\varepsilon=
\begin{cases}
\max\{0,t+v-b\},& A\ge b-v,\\
0,& A<b-v.
\end{cases}
\end{equation}
In particular, $0\le\varepsilon\le t-1$, and $\varepsilon=0$ when $t=1$.
\end{proposition}
\begin{proof}
For integers $0\le a\le A$, let
$g(a)=a+t\lfloor(L-a)/b\rfloor$.
Since $L-A=bu+v$, we have $g(A)=A+tu$.
Writing $a=A-z$, where $0\le z\le A$, gives
\[
g(A-z)-g(A)
=-z+t\left\lfloor\frac{v+z}{b}\right\rfloor.
\]
Set $j=\frac{v+z}{b}$. If $j<1$, the floor is zero, the above gives $-z \le 0$.
If $j\ge1$, then $z\ge jb-v$, and hence
\[
g(A-z)-g(A)
\le v-j(b-t)
\le v-(b-t)=t+v-b,
\]
where the second inequality uses $b-t=r>0$.

If $A<b-v$, the floor is zero for every feasible $z$, so $a=A$
is optimal. If $A\ge b-v$, the choice $z=b-v$ is feasible and
attains the increment $t+v-b$; comparing it with $z=0$ gives
the stated $\varepsilon$. Thus
$\max_{0\le a\le A}g(a)=A+tu+\varepsilon$, which yields
$F_*=L-A-tu-\varepsilon$.
Finally, $0\le v\le b-1$ implies $0\le\varepsilon\le t-1$,
and therefore $\varepsilon=0$ when $t=1$.
\end{proof}
Hereafter, $F_*$ denotes the optimized bound in~\eqref{eq:closed-optimized}, and $F_{\mathrm{multi}}$ denotes the bound in Theorem~\ref{thm:theory-bound}.

\begin{corollary}[Asymptotic tradeoff for fixed local parameters]\label{cor:asymptotic}
Fix $r,\delta$, and consider a sequence of nontrivial codes in this model with $n\to\infty$, $k/n\to R$, and $d/n\to\Delta$. Let $\alpha=(\delta-1)/(r+\delta-1)$. Then
\begin{equation}\label{eq:asymptotic}
R\le (1-\alpha)^2-(1-\alpha)(2-\alpha)\Delta.
\end{equation}
In particular, if $R=0$, then $\Delta\le r/(2r+\delta-1)$.
\end{corollary}
\begin{proof}
Divide the bound in Theorem~\ref{thm:theory-bound} by $n$. The sizes of the local sets selected in the two rounds, normalized by $n$, tend to $\alpha(1-\Delta)$ and $\alpha[1-2\Delta-\alpha(1-\Delta)]$, respectively. Substitution gives~\eqref{eq:asymptotic}. The bounded integer correction in Proposition~\ref{prop:closed-optimized} does not affect this limit.
\end{proof}

\subsection{A Pure-Code Bound Allowing Arbitrary Overlaps}
\label{pure-subsection}
We first relate the local erasure tolerance to the global distance.
This preliminary observation does not require purity or disjoint
recovery blocks.

\begin{lemma}[Global distance and local erasure tolerance]
\label{pure-distance}
Every nontrivial $(r,\delta)$-qLRC under
Definition~\ref{def:prelim-rdelta} satisfies $d\ge\delta$,
whether or not it is pure.
\end{lemma}
\begin{proof}
Take any $F\subseteq[n]$ with $|F|\le\delta-1$.
The empty set is trivially correctable.
If $F$ is nonempty, choose $i\in F$ and let
$\Gamma=\Gamma_i$ be its designated recovery block.
Set $E=F\cap\Gamma$ and $W=F\setminus\Gamma$.
Since $|E|\le\delta-1$, there is a recovery channel
$\operatorname{Rec}_{\Gamma,E}:\mathcal B(\mathcal H_{\Gamma\setminus E})
\to\mathcal B(\mathcal H_\Gamma)$.
Taking the partial trace over $W\subseteq\Gamma^c$ in its recovery
identity gives
\[
 \begin{aligned}
 &(\operatorname{Rec}_{\Gamma,E}\otimes
   \operatorname{id}_{\Gamma^c\setminus W})(\operatorname{Tr}_F\tau)\\
 &\qquad=\operatorname{Tr}_W\!\left[
   (\operatorname{Rec}_{\Gamma,E}\otimes\operatorname{id}_{\Gamma^c})
   (\operatorname{Tr}_E\tau)\right]
 =\operatorname{Tr}_W\tau
 \end{aligned}
\]
for every $\tau\in\mathcal D(Q)$.
Each step corrects at least one erasure, and the remaining
erasure set still has size at most $\delta-1$.
Repeating the procedure corrects all erasures in $F$.
Hence $d\ge\delta$.
\end{proof}

For the rest of this subsection, we additionally assume that
$Q$ is pure. Relative to the tensor-product error basis fixed
in Section~\ref{sec:preliminaries}, this means that every
tensor-product basis error $E$ satisfies
\begin{equation}\label{pure-definition}
 PEP=0
 \qquad\text{whenever }0<\operatorname{wt}(E)<d.
\end{equation}
We next record the resulting form of the small marginals
of $\rho=P/K$.

\begin{lemma}[Maximally mixed marginals of pure codes]
\label{lem:pure-marginals}
Let $Q$ be a pure $((n,K,d))_q$ quantum code with
orthogonal projector $P$, and let $\rho=P/K$.
For every $T\subseteq[n]$ with $|T|\le d-1$,
\begin{equation}\label{pure-marginals}
 \rho_T=\frac{I_T}{q^{|T|}}.
\end{equation}
\end{lemma}
\begin{proof}
Let $\{F_\alpha\}$ be the tensor-product error basis on
$\mathcal H_T$, with $F_0=I_T$.
For $\alpha\ne0$, the extension
$E_\alpha=F_\alpha\otimes I_{T^c}$ has
$0<\operatorname{wt}(E_\alpha)\le |T|<d$.
Purity gives $PE_\alpha P=0$ and hence
$PE_\alpha^\dagger P=0$. Therefore,
\[
\begin{aligned}
 \operatorname{Tr}_{\mathcal H_T}
 (F_\alpha^\dagger\rho_T)
 &=\frac1K\operatorname{Tr}_{\mathcal H}
 (E_\alpha^\dagger P)\\
 &=\frac1K\operatorname{Tr}_{\mathcal H}
 (PE_\alpha^\dagger P)=0.
\end{aligned}
\]
Orthogonality gives
$\operatorname{Tr}_{\mathcal H_T}(F_\alpha^\dagger F_\beta)
=q^{|T|}\delta_{\alpha\beta}$, so
\[
 \rho_T=\frac1{q^{|T|}}\sum_\alpha
 \operatorname{Tr}_{\mathcal H_T}
 (F_\alpha^\dagger\rho_T)F_\alpha.
\]
All nonidentity terms vanish, while the coefficient inside
the sum for $F_0=I_T$ is
$\operatorname{Tr}_{\mathcal H_T}\rho_T=1$.
Thus $\rho_T=I_T/q^{|T|}$.
\end{proof}

The preceding marginal identity combines with local erasure
correctability to give decoupling from the entire system
outside a recovery block, including the reference system.

\begin{lemma}[Local decoupling from an external reference system]
\label{pure-local-decoupling}
Let $Q$ be a nontrivial pure $(r,\delta)$-qLRC, let
$\Gamma$ be a designated recovery block, and let
$E\subseteq\Gamma$ with $|E|\le\delta-1$.
In the purification~\eqref{eq:prelim-code-purification},
set $W=R_0\Gamma^c$. Then
\begin{equation}\label{pure-joint-reference}
 \rho_{EW}=\frac{I_E}{q^{|E|}}\otimes\rho_W.
\end{equation}
\end{lemma}
\begin{proof}
By Lemma~\ref{lem:theory-support}, the local support code
$Q_\Gamma$ corrects erasures on $E$.
The global state $|\Omega\rangle$ purifies
$\rho_\Gamma\in\mathcal D(Q_\Gamma)$, with reference
system $W$.
Applying Lemma~\ref{lem:prelim-decoupling} to $Q_\Gamma$,
with erased subsystem $E$ and remaining subsystem
$\Gamma\setminus E$, gives
\[
 \rho_{EW}=\rho_E\otimes\rho_W.
\]
By Lemma~\ref{pure-distance},
$|E|\le\delta-1\le d-1$.
Lemma~\ref{lem:pure-marginals} therefore gives
$\rho_E=I_E/q^{|E|}$, proving the claim.
\end{proof}

\begin{theorem}[Pure-code Singleton-like bound]
\label{pure-bound}
\label{thm:pure-overlap}
Let $Q$ be a pure $((n,K,d))_q$ $(r,\delta)$-qLRC with $K>1$. Write $k=\log_qK$, $t=\delta-1$, $b=r+t$, and $D=d-1$. Then
\begin{equation}\label{pure-budget}
 k\le n-2\max_{\substack{e\in\mathbb Z\\0\le e\le D}}
 \left\{e+t\left\lfloor\frac{n-e}{b}\right\rfloor\right\}.
\end{equation}
Writing $n-D=bM+v$ with $0\le v<b$, the right-hand side is exactly
\begin{equation}\label{pure-explicit}
 F_{\mathrm{pure}}
 :=n-2D-2tM-2\max\{0,v-r\}.
\end{equation}
No integrality assumption on $k$ is required.
\end{theorem}
\begin{proof}
Fix an integer $0\le e\le D$.
Since $b\ge2t+1$ by Lemma~\ref{lem:theory-blocksize},
Lemma~\ref{lem:theory-selection} with $R=[n]$ and budget $n-e$
gives $S\subseteq C$ with
$|S|=t\lfloor(n-e)/b\rfloor$ and $|C|\le n-e$.
Choose $V\subseteq[n]\setminus C$ of size $e$.
The selection order, preserved after shrinking the selected set,
partitions $S=\bigsqcup_{i=1}^{\ell}S_i$ with
\begin{equation}\label{pure-selection-order}
 S_i\subseteq\Gamma_i,\quad |S_i|\le t,\qquad
 W_i:=V\cup\bigcup_{j>i}S_j\subseteq\Gamma_i^c.
\end{equation}
By~\eqref{pure-joint-reference},
$\rho_{S_iW_i}=q^{-|S_i|}I_{S_i}\otimes\rho_{W_i}$.
Successively factoring out $S_1,\ldots,S_\ell$ and using
$\rho_V=I_V/q^e$ from~\eqref{pure-marginals} gives
\begin{equation}\label{pure-selected-marginal}
 \rho_{S\cup V}
 =\left(\bigotimes_{i=1}^{\ell}\frac{I_{S_i}}{q^{|S_i|}}\right)
   \otimes\frac{I_V}{q^e}
 =\frac{I_{S\cup V}}{q^{|S|+e}}.
\end{equation}
Moreover, $A=S\cup V$ is correctable by local recovery of $S$
followed by distance-based recovery of $V$.
Corollary~\ref{cor:prelim-correctable-entropy} yields
$k\le n-2|A|$.
Substituting $|A|=e+t\lfloor(n-e)/b\rfloor$ and optimizing
over $e$ proves~\eqref{pure-budget}.

For the closed form, write $n-D=bM+v$ and $e=D-z$.
Relative to $e=D$, the objective increases by
$-z+t\lfloor(v+z)/b\rfloor$.
If the floor is zero, this is nonpositive; if it equals
$j\ge1$, then $z\ge jb-v$, so the increase is at most
$v-jr\le v-r$.
Thus the maximum increase is at most $\max\{0,v-r\}$.
For $v\le r$, $z=0$ attains it.
For $v>r$, $z=b-v$ attains it and is feasible because
$0<b-v<t\le D$ by Lemma~\ref{pure-distance}.
This proves~\eqref{pure-explicit}.
\end{proof}

\begin{corollary}[Equivalent implicit form]
\label{pure-implicit-corollary}
The bound in Theorem~\ref{pure-bound} is equivalent to
\begin{equation}\label{pure-implicit}
 k+2d+2(\delta-1)
 \left(\left\lceil\frac{n+k}{2r}\right\rceil-1\right)
 \le n+2.
\end{equation}
\end{corollary}
\begin{proof}
Recall that $t=\delta-1$, $b=r+t$, $D=d-1$, and
$n-D=bM+v$ with $0\le v<b$.
Set $x=(n+k)/2\in\mathbb R$.
Equation~\eqref{pure-implicit} is equivalent to
\[
 f(x):=x+t\bigl(\lceil x/r\rceil-1\bigr)\le bM+v.
\]
The function $f$ is increasing. We show that its largest feasible
argument is $x_*=rM+\min\{v,r\}$.

Since $\lceil x_*/r\rceil\le M+1$, we have
$f(x_*)\le bM+\min\{v,r\}\le bM+v$, so $x_*$ is feasible.
If $v\le r$ and $x>x_*=rM+v$, then
$\lceil x/r\rceil\ge M+1$, giving
$f(x)\ge x+tM>bM+v$.
If $v>r$ and $x>x_*=r(M+1)$, then
$\lceil x/r\rceil\ge M+2$, and hence
\[
 f(x)\ge x+t(M+1)>b(M+1)>bM+v.
\]
Thus no $x>x_*$ is feasible. Monotonicity of $f$ now gives
$k\le2[rM+\min\{v,r\}]-n$, which is precisely
\eqref{pure-explicit}. No integrality assumption on $k$ is used.
\end{proof}

For pure codes obtained from Euclidean or Hermitian dual-containing
classical constructions, \eqref{pure-implicit} coincides with the
bound in~\cite[Theorem~31 and Remark~34]{Galindo2026Locality}.
A detailed comparison and the extension to arbitrary pure quantum
subspaces are discussed in Section~\ref{subsec:pure-comparison}.

\section{Equality Conditions and Bounds at Minimal Locality}\label{sec:equality-structure}
We first study the local structure forced by equality in the general
entropy bound. We then use the $\mathbb F_q$-linear stabilizer structure
to derive a decomposition theorem and a sharp dimension bound for
$r=\delta$, without requiring equality in that entropy bound.

\subsection{Equality Constraints and Isolated Local MDS Blocks}
By retaining the mutual-information terms in the quantum Singleton
entropy argument~\cite{Preskill2016Shannon}, we obtain
the exact identity~\eqref{eq:equality-deficit}, which relates the code
dimension parameter $k$ to $S(B)$, $I(A_1:B)$, and $I(A_2:B)$.
We first characterize equality in $k\le|B|$ and then combine these
conditions with local multiple-erasure recovery to derive necessary
conditions for attaining our parameter bounds.
\begin{proposition}[Equality conditions for a fixed correctable partition]
\label{prop:equality-partition}
Let $[n]=A_1\sqcup A_2\sqcup B$, where $A_1,A_2$ are correctable erasure sets for a quantum code $Q$ of dimension $K=q^k$. Write $\rho=P_Q/K$. Then
\begin{equation}\label{eq:equality-deficit}
k=S(B)-\tfrac12\bigl(I(A_1:B)+I(A_2:B)\bigr).
\end{equation}
Consequently, $k=|B|$ if and only if
\begin{equation}\label{eq:equality-products}
\rho_B=I_B/q^{|B|},\qquad
\rho_{A_iB}=\rho_{A_i}\otimes I_B/q^{|B|}\quad(i=1,2).
\end{equation}
Moreover, when equality holds, for every $U\subseteq[n]$, setting $X=U\cap A_1$, $Y=U\cap A_2$, and $Z=U\cap B$ gives
\begin{equation}\label{eq:equality-subset-entropy}
S(U)\ge |Z|+|S(X)-S(Y)|\ge |Z|.
\end{equation}
\end{proposition}
\begin{proof}
Correctability and a purification with reference system $R_0$ give $k+S(A_1)=S(A_2B)$ and $k+S(A_2)=S(A_1B)$. Adding yields~\eqref{eq:equality-deficit}. Since mutual information is nonnegative and $S(B)\le|B|$, equality holds if and only if~\eqref{eq:equality-products} holds. Under equality, $\rho_{XZ}=\rho_X\otimes I_Z/q^{|Z|}$, and similarly for $YZ$. Applying the Araki--Lieb inequality to the partitions $XZ:Y$ and $YZ:X$ gives
\[
S(U)\ge S(X)+|Z|-S(Y),\qquad
S(U)\ge S(Y)+|Z|-S(X),
\]
which proves~\eqref{eq:equality-subset-entropy}.
\end{proof}

\begin{theorem}[Constraints under equality]
\label{thm:equality-local-intersections}
Under the assumptions of Proposition~\ref{prop:equality-partition}, suppose further that $k=|B|$ and that $Q$ is an $(r,\delta)$-qLRC. Write $t=\delta-1$. For any designated recovery block $\Gamma$ meeting $B$, set $X=\Gamma\cap A_1$, $Y=\Gamma\cap A_2$, and $Z=\Gamma\cap B$. Then
\begin{equation}\label{eq:equality-intersections}
|X|\ge t,\qquad |Y|\ge t,\qquad
1\le |Z|\le |\Gamma|-2t\le r-t.
\end{equation}
Moreover,
\begin{equation}\label{eq:equality-local-rank}
|Z|+|S(X)-S(Y)|\le S(\Gamma)
\le\log_q\dim Q_\Gamma\le |\Gamma|-2t.
\end{equation}
If $|X|=|Y|=t$, then $\rho_\Gamma$ is the maximally mixed state on a quantum MDS code with parameters $((2t+|Z|,q^{|Z|},t+1))_q$.
\end{theorem}
\begin{proof}
By Lemma~\ref{lem:prelim-decoupling}, if the local support
code $Q_\Gamma$ corrects erasures on $E$, then $E$ is
uncorrelated with the reference system in any purification
of a state supported on $Q_\Gamma$.

Suppose $|X|<t$. Choose $j\in Z$ and write $Z'=Z\setminus\{j\}$. By Lemma~\ref{lem:theory-support}, the local support code $Q_\Gamma$ corrects erasures on $E=X\cup\{j\}$. Purify $\rho_\Gamma$ with reference system $W$. Decoupling of $E$ from $W$, together with equality of complementary entropies for a pure state, gives
\[
S(\Gamma)=S(YZ')-S(Xj)
=S(Y)-S(X)+|Z|-2,
\]
where the second equality uses~\eqref{eq:equality-products}. However, Araki--Lieb also gives $S(\Gamma)\ge S(YZ)-S(X)=S(Y)-S(X)+|Z|$, a contradiction. Thus $|X|\ge t$. Interchanging $A_1,A_2$ gives $|Y|\ge t$, proving~\eqref{eq:equality-intersections}.

Equation~\eqref{eq:equality-subset-entropy} gives the first inequality in~\eqref{eq:equality-local-rank} and shows that $Q_\Gamma$ has dimension greater than one. The remaining inequalities follow from the entropy--rank bound and the quantum Singleton bound for the local support code, which corrects any set of at most $t$ erasures. If $|X|=|Y|=t$, all inequalities between $|Z|$ and $\log_q\dim Q_\Gamma$ are equalities. Hence $\dim Q_\Gamma=q^{|Z|}$ and $\rho_\Gamma=P_{Q_\Gamma}/q^{|Z|}$. The local distance is at least $t+1$, and the Singleton bound rules out a larger distance, giving the stated MDS parameters.
\end{proof}

\par\smallskip\noindent\textbf{Example: a four-qubit MDS block.}
Let $Q$ be the binary stabilizer code generated by $X^{\otimes4}$ and $Z^{\otimes4}$, with parameters $\llbracket4,2,2\rrbracket_2$. Take $A_1=\{1\}$, $A_2=\{2\}$, and $B=\{3,4\}$, and designate the same recovery block $\Gamma=[4]$ for every coordinate. Both $A_i$ are correctable, $k=|B|=2$, and $(r,\delta)=(3,2)$, so $t=1$. Here $(|X|,|Y|,|Z|)=(1,1,2)$, while $S(X)=S(Y)=1$ and $S(\Gamma)=2$. Thus every inequality in~\eqref{eq:equality-local-rank} is an equality:
\[
|Z|+|S(X)-S(Y)|=S(\Gamma)
=\log_2\dim Q_\Gamma=|\Gamma|-2t=2.
\]
This illustrates the MDS conclusion when both intersections have their minimum sizes: the local code has dimension exactly $2^{|Z|}=4$ and attains the quantum Singleton bound.

\begin{theorem}[Isolation at the minimum recovery-block length]
\label{thm:equality-isolation}
Under the assumptions of Theorem~\ref{thm:equality-local-intersections}, suppose further that $r=\delta$. Every designated recovery block $\Gamma$ meeting $B$ satisfies
\[
|\Gamma|=2\delta-1,\qquad |\Gamma\cap B|=1,
\]
and $\rho_\Gamma$ is the maximally mixed state on an $((2\delta-1,q,\delta))_q$ quantum MDS code. Moreover, for any designated recovery block $\Gamma'$,
\begin{equation}\label{eq:equality-isolation}
\Gamma'=\Gamma\quad\text{or}\quad\Gamma'\cap\Gamma=\varnothing.
\end{equation}
Consequently, the designated recovery blocks include at least $|B|=k$ distinct, pairwise disjoint isolated blocks, each of size $2\delta-1$.
\end{theorem}
\begin{proof}
Let $t=\delta-1$. The first part follows directly from Theorem~\ref{thm:equality-local-intersections} and $|\Gamma|\le2t+1$. We next prove the entropy property needed for isolation, without assuming a stabilizer representation of the local code.

Entropy formulas for quantum MDS codes are known; see~\cite{Sun2025EntropyMDS}. For completeness, we derive the special case needed here. For the maximally mixed state on an $((2t+1,q,t+1))_q$ MDS code,
\begin{equation}\label{eq:local-mds-entropy-profile}
S(T)=\min\{|T|,2t+2-|T|\}\qquad(T\subseteq\Gamma).
\end{equation}
Indeed, choose any $(t+1)$-element subset $T$ and any $j\in T$. Both $t$-element sets $T\setminus\{j\}$ and $\Gamma\setminus T$ are correctable. The equality conditions in Proposition~\ref{prop:equality-partition} give $\rho_T=\rho_{T\setminus\{j\}}\otimes I_j/q$. This holds for every $j\in T$, so $\rho_T=I_T/q^{t+1}$, and all smaller marginals are maximally mixed as well. If $|T|>t+1$, its complement has size at most $t$ and is decoupled from the reference system, whose entropy is one. Purification then gives $S(T)=1+|\Gamma\setminus T|$, proving the formula.

Suppose that a designated block $\Gamma'\ne\Gamma$ meets $\Gamma$. Its local support code has dimension at most $q$: when the dimension exceeds one, this follows from the quantum Singleton bound and $|\Gamma'|\le2t+1$; when the dimension is one, it is immediate. Hence $S(\Gamma')\le1$. Set $H=\Gamma\cap\Gamma'$ and $U=\Gamma\cup\Gamma'$. Then $H$ is a nonempty proper subset of $\Gamma$. Strong subadditivity and~\eqref{eq:equality-subset-entropy} imply
\[
S(H)+1\le S(H)+S(U)
\le S(\Gamma)+S(\Gamma')\le2.
\]
By~\eqref{eq:local-mds-entropy-profile}, this is possible only when $H=\{j\}$ is a singleton. But local recovery of the erasure at $j$ within $\Gamma'$ implies that, in a purification of the global state, $j$ is decoupled from the system outside $\Gamma'$. Since $\Gamma\setminus\{j\}$ lies entirely outside $\Gamma'$, we must have $I(j:\Gamma\setminus\{j\})=0$. On the other hand, the MDS entropy formula gives
\[
I(j:\Gamma\setminus\{j\})=1+2-1=2,
\]
a contradiction. This proves~\eqref{eq:equality-isolation}. Finally, every coordinate in $B$ belongs to a designated block, and each such block contains exactly one coordinate of $B$.
\end{proof}

\par\smallskip\noindent\textbf{Example: two isolated three-coordinate blocks.}
Let $C$ be the $\llbracket3,1,2\rrbracket_3$ code with logical basis
\[
|a_L\rangle=\frac1{\sqrt3}
\sum_{x\in\mathbb F_3}|x,x+a,x+2a\rangle,
\qquad a\in\mathbb F_3.
\]
Take $Q=C\otimes C$ with designated recovery blocks $\Gamma_1=\{1,2,3\}$ and $\Gamma_2=\{4,5,6\}$. Then $Q$ has parameters $\llbracket6,2,2\rrbracket_3$ and $(r,\delta)=(2,2)$. The partition $A_1=\{1,4\},\; A_2=\{2,5\},\; B=\{3,6\}$ satisfies $k=|B|=2$. Each $A_i$ erases one coordinate in each of two independently recoverable blocks and is therefore correctable. Each $\Gamma_i$ has length three and contains exactly one coordinate of $B$, as the theorem asserts. Isolation also excludes any other distinct designated recovery block meeting either triple. For example, $\{1,2,4\}$ cannot be a recovery block: its reduced code state is $I_{124}/27$, so its local support is the entire space $(\mathbb C^3)^{\otimes3}$, which cannot correct a single erasure.

\begin{remark}[Structural obstructions to equality]
\label{cor:equality-connected-obstruction}
If $r=\delta$ and $Q$ attains the optimized bound $F_*>0$, the optimized partition satisfies $k=|B|=F_*$. Theorem~\ref{thm:equality-isolation} gives $F_*$ isolated blocks. A necessary condition for equality is therefore
{
\begin{equation}\label{eq:equality-numerical-obstruction}
(2\delta-1)F_*\le n.
\end{equation}
}
Thus equality is impossible if this condition fails, or if the designated recovery-block cover is connected through chains of intersecting blocks and $n>2\delta-1$. When equality is impossible, integer $k$ satisfies $k\le F_*-1$; in general, $K\le q^{F_*}-1$.
\end{remark}
These conclusions isolate only the recovery blocks meeting $B$; they do not decompose an arbitrary quantum code. The next theorem does not require equality in the global bound, but applies only to $\mathbb F_q$-linear stabilizer codes.

\subsection{Linear Stabilizer Structure and a Tighter Bound at Minimal Locality}
Here linearity means that, after phases are removed, the stabilizer labels form an $\mathbb F_q$-linear self-orthogonal space $S\subseteq\mathbb F_q^{2n}$ under the ordinary symplectic form. When $q=p^m$, this is stronger than $\mathbb F_p$-additivity alone. If an $\mathbb F_q$-linear label space is trace-symplectically self-orthogonal, applying the condition to scalar multiples by arbitrary field elements implies ordinary symplectic self-orthogonality. Write $\perp_s$ for the dual with respect to this form. For a coordinate set $J$, let $V_J$ be the symplectic coordinate subspace supported on $J$. For any label subspace $L$, write $L[J]=L\cap V_J$; in particular, $S[J]=S\cap V_J$. The reduced state of the normalized code projector is again the normalized projector of a local stabilizer code, whose support on $J$ has dimension $q^{|J|-\dim_{\mathbb F_q}S[J]}$.

\begin{theorem}[Structure and a tight bound when $r=\delta$]
\label{thm:minimal-locality-stabilizer}
Let $Q$ be an $\mathbb F_q$-linear stabilizer $(\delta,\delta)$-qLRC with parameters $\llbracket n,k,d\rrbracket_q$, where $k\ge1$, $\delta\ge2$, and $d$ is the exact minimum distance, not merely a distance lower bound. Set $t=\delta-1$ and $b=2\delta-1$. There exists a coordinate set $U$ such that
\[
Q=|\phi\rangle_U\otimes Q',
\]
where $|\phi\rangle_U$ is a fixed pure stabilizer state, and the remaining coordinates partition into $h$ disjoint blocks of length $b$. On each block, the local support of $Q'$ is a quantum MDS stabilizer code $\llbracket b,1,\delta\rrbracket_q$. Using these local codes as encoders, $Q'$ is their concatenation with an $\mathbb F_q$-linear outer stabilizer code $\llbracket h,k,d_{\mathrm{out}}\rrbracket_q$, and
\[
d=\delta d_{\mathrm{out}}.
\]
In particular,
\begin{equation}\label{eq:minimal-locality-sharp}
\delta\mid d,
\qquad
k\le h-2\left(\frac d\delta-1\right)
\le\left\lfloor\frac{n}{2\delta-1}\right\rfloor-2\frac d\delta+2.
\end{equation}
Conversely, let $n=h(2\delta-1)+z$, with $0\le z<2\delta-1$, $d_{\mathrm{out}}\ge1$, and $h\ge2d_{\mathrm{out}}-1$. For every prime power $q\ge\max\{2\delta-1,h\}$, such a code exists with exact parameters
\[
\llbracket n,h-2d_{\mathrm{out}}+2,\delta d_{\mathrm{out}}\rrbracket_q,
\]
attaining the final bound in~\eqref{eq:minimal-locality-sharp}.
\end{theorem}
\begin{proof}[Proof outline]
The Singleton bound for local support codes forces each local support code of dimension greater than one to be a $\llbracket2\delta-1,1,\delta\rrbracket_q$ MDS code. Symplectic projection properties of its stabilizer imply that two distinct such blocks are either disjoint or have a union supporting a fixed pure state. After a maximal fixed pure-state coordinate factor is removed, the remaining coordinates therefore partition into local MDS blocks. Quotienting by the direct sum of their stabilizers yields the outer code. Every nonzero logical Pauli class of an inner code has a representative of weight exactly $\delta$, giving both inequalities needed for $d=\delta d_{\mathrm{out}}$. The outer-code Singleton bound gives the stated bound, which is attained by nested Reed--Solomon CSS codes. Appendix~\ref{app:minimal-locality-proof} provides the complete intersection, factorization, and logical-coset arguments.
\end{proof}

\section{Comparison with Existing Bounds and Their Scope}\label{sec:comparison}
The preceding two sections establish a general subspace bound, a pure-code bound, and a linear stabilizer bound at minimal locality. We now use common notation to compare their numerical strength, assumptions, and relation to existing results.

\subsection{Relevant Existing qLRC Bounds}\label{subsec:known-bounds}
This subsection collects upper bounds relevant to our model and distinguishes results that require additional algebraic or recovery structure. Throughout this comparison, we consider nontrivial codes ($K>1$). $d$ denotes the actual quantum distance, $k=\log_q\dim Q$, $D=d-1$, $t=\delta-1$, and $b=r+t$. The locality convention in~\cite{Golowich2023LRC} counts the erased coordinate, whereas our single-erasure locality $r$ counts at most $r$ helper coordinates. We denote classical dimensions by $m$ or $m_i$ and explicitly write the classical distance as $d_C$, to avoid applying an inequality involving classical distance to a potentially larger impure quantum distance.

\subsubsection{General Single-Erasure Bounds and CSS Bounds}
The quantum Singleton bound is $k\le n-2D$\cite{Rains1999Nonbinary}. For general quantum subspace qLRCs with recovery blocks of size at most $b$, Golowich and Guruswami~\cite[full version, Theorem~35]{Golowich2023LRC} proved
\begin{equation}\label{eq:known-gg}
 k\le n-2D-u-\left\lfloor\frac{n-2D-u}{b}\right\rfloor,
 \qquad u=\left\lfloor\frac{n-D}{b}\right\rfloor.
\end{equation}
When $\delta>2$, this substitution uses only the single-erasure recovery capability. If the recovery blocks form an equal-sized partition, their Theorem~36 gives
\begin{equation}\label{eq:known-gg-disjoint}
 k\le\left(1-\frac2b\right)n-2D+2\left\lceil\frac{D}{b-1}\right\rceil.
\end{equation}
Neither result requires purity.

Suppose a CSS code is constructed from two classical codes of dimensions $m_1,m_2$ satisfying the common-support locality condition in~\cite[Proposition~2]{Luo2023CSS}. Then $k=m_1+m_2-n$. For single-erasure locality $r$, Luo et al.~\cite[Corollaries~5 and~6]{Luo2023CSS} established
\begin{equation}\label{eq:known-css-singleton}
 2d\le n-k-2\left\lceil\frac{k}{r}\right\rceil+4,
\end{equation}
and the Cadambe--Mazumdar (CM)-type bound
\begin{equation}\label{eq:known-css-cm}
 2k\le\min_{\ell_1,\ell_2}
 \sum_{i=1}^{2}\left\{\ell_i r+
 k_{\mathrm{opt}}^{(q)}\bigl(m_i-\ell_i(r+1),d\bigr)\right\}.
\end{equation}
Here $k_{\mathrm{opt}}^{(q)}(N,d)$ is the largest dimension of a linear $q$-ary code of length $N$ and distance at least $d$. The nonnegative integer shortening indices range only over values for which the remaining lengths are valid. Their Theorem~4 also gives transfer inequalities involving the optimal classical distance and length functions. These results allow impure CSS codes satisfying the construction assumptions. Equations~\eqref{eq:known-css-singleton} and~\eqref{eq:known-gg} are not equivalent formulas: their Theorem~9 shows that a pure CSS code attaining the former also attains the latter. Attainment implications must therefore be distinguished from pointwise equality of the bounds.

\subsubsection{Multiple Erasures, Disjoint Blocks, and Hierarchical Recovery}
Let $C$ be a Euclidean or Hermitian dual-containing classical $(r,\delta)$-LRC of dimension $m=(n+k)/2$. Galindo et al.~\cite[Theorem~31]{Galindo2026Locality} proved
\begin{equation}\label{eq:known-galindo}
 k+2d_C+2t\left(\left\lceil\frac{n+k}{2r}\right\rceil-1\right)\le n+2.
\end{equation}
The resulting quantum code has distance $d\ge d_C$. Purity permits the substitution $d_C=d$; their Remark~34 explicitly notes that this replacement is not generally valid in the impure case. The counterexamples in~\cite{Galindo2026Impure} show that this distinction is essential.

For a partition into $s=n/b$ equal-sized recovery blocks, Guruswami, Kshirsagar, and Trivedi~\cite[v2, Theorem~2.10]{Guruswami2026Hierarchical} give the general multiple-erasure bound
\begin{equation}\label{eq:known-multi-disjoint}
 k\le n-2D-2t\left(s-\left\lceil\frac{D}{r}\right\rceil\right).
\end{equation}
This also appears as Eq.~(7) in~\cite{StylianouBoche2026} and equals~\eqref{eq:known-gg-disjoint} when $\delta=2$. The version matters here: Theorem~2.10 in v1 of the hierarchical-code paper gives~\eqref{eq:known-gg-disjoint} with $b=r+\delta-1$.

For pure quantum subspaces with these disjoint blocks, Stylianou and Boche~\cite[Theorem~4]{StylianouBoche2026} obtain
\begin{equation}\label{eq:known-pure-disjoint}
 k\le n-2D-2t\left\lfloor\frac{n-D}{b}\right\rfloor.
\end{equation}
Their Theorems~5 and~6 give a blockwise linear programming (LP) bound and an explicit Singleton-like consequence. Writing $d=zb+\epsilon$ with $1\le\epsilon\le b$, the latter is
\begin{equation}\label{eq:known-lp-explicit}
 k\le r(s-2z)-st-2\max\{0,\epsilon-\delta\}.
\end{equation}
More generally, let $\nu,\mu\in\{0,\ldots,b\}^s$ and use the product basis $K_\nu(\mu)=\prod_jK_{\nu_j}^{(b,q^2)}(\mu_j)$. Consider a polynomial of degree at most $b$ in each variable with blockwise Krawtchouk expansion $f=\sum_{\nu}f_{\nu}K_{\nu}$. If $f_{\nu}\ge0$, $f_0>0$, and $f(\mu)\le0$ at every grid point satisfying $\sum_{j:\mu_j\ge\delta}\mu_j\ge d$, their LP theorem gives
\begin{equation}\label{eq:known-lp}
 \dim Q\le q^{-n}\frac{f(0)}{f_{0}}.
\end{equation}
Their block-pattern purity condition is discussed in Remark~\ref{pure-block-pattern}. When $n=sb$, the explicit exponent in~\eqref{eq:known-lp-explicit} equals the optimized expression in Theorem~\ref{pure-bound}, which extends it to arbitrary pure quantum subspaces with possibly overlapping recovery blocks.

For hierarchical recovery, let $r_1\ge\cdots\ge r_h$, $\delta_1\ge\cdots\ge\delta_h\ge2$, and $\delta_{h+1}=1$. If a dual-containing classical $h$-level HLRC has parameters $(r_j,\delta_j)$ and $d(C^\perp)\ge\delta_1$, then $\mathrm{CSS}(C,C)$ satisfies~\cite[v2, Proposition~2.11]{Guruswami2026Hierarchical}
\begin{equation}\label{eq:known-hierarchy}
 \begin{split}
 k+2d_C\le n+2-2\sum_{j=1}^{h}
 \left(\left\lceil\frac{n+k}{2r_j}\right\rceil-1\right)
 (\delta_j-\delta_{j+1}).
 \end{split}
\end{equation}
This expression uses the classical distance $d_C$; substitution of $d$ is valid only when the two distances coincide. The hierarchical model requires nested recovery levels.

\subsubsection{Alphabet-Dependent Bounds for Pure Hermitian Constructions}
Li et al.~\cite[Theorem~3]{Li2025Pure} give Griesmer-type, Plotkin-type, and sphere-packing bounds for single-erasure pure Hermitian qLRCs; Kumar and Bandi~\cite[Theorem~3]{KumarBandi2026} give the corresponding $(r,\delta)$ versions. Use the common notation $m=(n+k)/2\in\mathbb Z$, $T=\lceil m/r\rceil-1$, and $T_s=\min\{T,\lfloor(n-1)/b\rfloor\}$, and set
\[
 V_{\eta}(N,e)=\sum_{j=0}^{\min(e,N)}\binom Nj(\eta-1)^j,
 \qquad e=\lfloor D/2\rfloor.
\]
Here $\eta\ge2$ is an integer alphabet size. For the pure Hermitian codes specified in those works,
\begin{align}
 n&\ge\max_{0\le\ell\le T}
 \left\{\ell b+\sum_{j=0}^{m-\ell r-1}\left\lceil\frac d{q^{2j}}\right\rceil\right\},
 \label{eq:known-griesmer}\\
 d&\le\min_{0\le\ell\le T}
 \frac{q^{2(m-\ell r)-2}(q^2-1)(n-\ell b)}{q^{2(m-\ell r)}-1},
 \label{eq:known-plotkin}\\
 k&\le n-2\max_{0\le\ell\le T_s}
 \{\ell t+\log_{q^2}V_{q^2}(n-\ell b,e)\}.
 \label{eq:known-packing}
\end{align}
The extrema are taken over integer indices. The common range of residual-code parameters above suffices to state all three types of bounds; the single-erasure sphere-packing theorem in~\cite{Li2025Pure} also gives the larger range $0\le\ell\le\lfloor(n-1)/(r+1)\rfloor$. These alphabet-dependent results rely on classical Hermitian constituent codes and cannot be asserted to hold directly for arbitrary pure quantum subspaces. Their asymptotic refinements follow by normalizing the finite-length bounds and retain the corresponding field-size dependence.

The general bound in Section~\ref{ch:bound} applies to quantum subspaces allowing both impurity and arbitrary overlaps; the pure-code bound extends the scope of~\eqref{eq:known-galindo} and~\eqref{eq:known-lp-explicit}. Section~\ref{sec:constructions} gives corresponding attaining examples.

\begin{table}[htbp]
\centering\footnotesize
\caption{Existing upper bounds and their additional assumptions.}
\label{tab:known-bound-inventory}
\begin{tabularx}{\linewidth}{@{}YYY@{}}
\toprule
Bound and reference & Assumptions beyond the basic parameters & Location in this paper\\
\midrule
Golowich--Guruswami Singleton bound~\cite{Golowich2023LRC}
& General subspaces; single-erasure recovery
& \eqref{eq:known-gg}\\
Golowich--Guruswami partition bound~\cite{Golowich2023LRC}
& Equal-sized disjoint recovery blocks
& \eqref{eq:known-gg-disjoint}\\
CSS Singleton and CM bounds of Luo et al.~\cite{Luo2023CSS}
& CSS construction with common supports
& \eqref{eq:known-css-singleton}--\eqref{eq:known-css-cm}\\
Classical-to-quantum multiple-erasure bound of Galindo et al.~\cite{Galindo2026Locality}
& Dual-containing constituent code; distance $d_C$
& \eqref{eq:known-galindo}\\
Guruswami--Kshirsagar--Trivedi multiple-erasure partition bound~\cite{Guruswami2026Hierarchical}
& Equal-sized disjoint recovery blocks; no purity assumption
& \eqref{eq:known-multi-disjoint}\\
Stylianou--Boche pure-code partition and LP bounds~\cite{StylianouBoche2026}
& Pure subspaces; equal-sized partition
& \eqref{eq:known-pure-disjoint}--\eqref{eq:known-lp}\\
Guruswami--Kshirsagar--Trivedi hierarchical Singleton bound~\cite{Guruswami2026Hierarchical}
& Dual-containing classical HLRC
& \eqref{eq:known-hierarchy}\\
Griesmer, Plotkin, and sphere-packing bounds of Li et al. and Kumar--Bandi~\cite{Li2025Pure,KumarBandi2026}
& Pure Hermitian constructions
& \eqref{eq:known-griesmer}--\eqref{eq:known-packing}\\
\bottomrule
\end{tabularx}
\end{table}

\subsection{Scope of the Parameter Bounds}\label{subsec:bound-scope}
Table~\ref{tab:scope-audit} summarizes the assumptions of the bounds above and our results. Numerical comparisons between pure-code and general bounds are made only on their common domain of pure codes. Denote the right-hand side of~\eqref{eq:known-multi-disjoint} by $F_{\mathrm{dis}}$. That bound requires a disjoint partition into recovery blocks of size $b=r+\delta-1$, but does not require purity. Neither our general bound nor our pure-code bound assumes such a partition. When $\delta=2$, it reduces to the Golowich--Guruswami disjoint-block bound~\eqref{eq:known-gg-disjoint}.
\begin{table}[htbp]
\centering\small
\caption{Relevant parameter bounds and their assumptions.}\label{tab:scope-audit}
\begin{tabularx}{\linewidth}{@{}>{\raggedright\arraybackslash}p{.28\linewidth}YY@{}}
\toprule Result & Code class & Recovery-block assumptions\\\midrule
Golowich--Guruswami~\cite{Golowich2023LRC} & General subspaces & Single erasure; arbitrary overlaps\\
Galindo et al.~\cite{Galindo2026Locality} & Euclidean/Hermitian constructions with the specified distance assumptions & Multiple erasures\\
Guruswami et al.~\cite{Guruswami2026Hierarchical} & General subspaces & Equal-sized disjoint blocks\\
Stylianou--Boche~\cite{StylianouBoche2026} & Pure subspaces, formulated through block patterns & Disjoint blocks\\
Theorem~\ref{thm:theory-bound}, Corollary~\ref{cor:theory-optimization} & General subspaces & Arbitrary overlaps\\
Theorem~\ref{thm:pure-overlap} & Pure subspaces in the usual sense & Arbitrary overlaps\\
Theorem~\ref{thm:minimal-locality-stabilizer} & $\mathbb F_q$-linear stabilizer codes & $r=\delta$; no assumed partition\\\bottomrule
\end{tabularx}
\end{table}

\subsection{Pointwise Comparison with the Golowich--Guruswami Bound}
To align the two locality conventions, set the maximum recovery-block size in the Golowich--Guruswami bound to $r+\delta-1$. This substitution uses only single-erasure recovery within each block; the block sizes need not be equal. Write $x=\left\lfloor\frac{n-d+1}{r+\delta-1}\right\rfloor$, and express the single-erasure benchmark~\cite[full version, Theorem~35]{Golowich2023LRC} as
\begin{equation}\label{eq:comparison-gg}
F_{\mathrm{GG}}=n-2d+2-x-
\left\lfloor\frac{n-2d+2-x}{r+\delta-1}\right\rfloor.
\end{equation}
Denote the right-hand side of Theorem~\ref{thm:theory-bound} by
\begin{equation}\label{eq:comparison-multi}
F_{\mathrm{multi}}=n-2d+2-(\delta-1)x-
(\delta-1)\left\lfloor
\frac{n-2d+2-(\delta-1)x}{r+\delta-1}\right\rfloor.
\end{equation}
\begin{proposition}[Comparison with the single-erasure benchmark]\label{prop:comparison-gg}
For parameters admitting a quantum $(r,\delta)$-qLRC with $K>1$, we have $F_{\mathrm{multi}}\le F_{\mathrm{GG}}$. The two are equal when $\delta=2$. If
\[
(\delta-2)\left\lfloor\frac{n-d+1}{r+\delta-1}\right\rfloor\ge2,
\]
then $F_{\mathrm{multi}}<F_{\mathrm{GG}}$.
\end{proposition}
\begin{proof}
When $\delta=2$, equations~\eqref{eq:comparison-gg} and~\eqref{eq:comparison-multi} agree term by term. Suppose henceforth that $\delta>2$. Theorem~\ref{thm:theory-bound} ensures $n-2d+2-(\delta-1)x\ge0$, so write
\[
n-2d+2-(\delta-1)x=(r+\delta-1)u+v,
\qquad u\ge0,\quad 0\le v<r+\delta-1,
\]
where $u,v$ are integers. Let $h=(\delta-2)x$. Then
\[
\left\lfloor\frac{n-2d+2-x}{r+\delta-1}\right\rfloor
=u+\left\lfloor\frac{v+h}{r+\delta-1}\right\rfloor.
\]
Substitution into the two bounds gives
\begin{equation}\label{eq:theory-gap}
F_{\mathrm{GG}}-F_{\mathrm{multi}}
=h+(\delta-2)u-
\left\lfloor\frac{v+h}{r+\delta-1}\right\rfloor.
\end{equation}
Since $h$ is a nonnegative integer and $v+h\le r+\delta-2+h<(h+1)(r+\delta-1)$, the floor term is at most $h$, so the difference is nonnegative. If $h\ge2$, then $v+h\le r+\delta-2+h<h(r+\delta-1)$, and the floor term is at most $h-1$, making the difference at least $1$.
\end{proof}

\subsection{The Stronger Pure-Code Bound and Extensions of Existing Bounds}
\label{subsec:pure-comparison}
Using the notation of Theorem~\ref{pure-bound}, let $D=d-1$, $t=\delta-1$, $b=r+t$, and $M=\lfloor(n-D)/b\rfloor$. On the class of pure codes,
\begin{equation}\label{pure-general-comparison}
 F_{\mathrm{pure}}\le n-2D-2tM\le F_*.
\end{equation}
\begin{proof}
To compare the bounds, set $A_0=tM$ and $L=n-2D$. For every integer $0\le a\le A_0$, the inequality $L-a\le n-D$ implies
\[
 a+t\left\lfloor\frac{L-a}{b}\right\rfloor\le2A_0.
\]
Equation~\eqref{eq:theory-opt} then gives the second inequality in~\eqref{pure-general-comparison}; the first follows directly from~\eqref{pure-explicit}.
\end{proof}
The inequality $F_{\mathrm{pure}}\le F_*$ in~\eqref{pure-general-comparison} can hold with equality or be strict. The glued GRS construction in Section~\ref{subsec:glued-grs} realizes both $F_{\mathrm{pure}}<F_*$ and $k=F_{\mathrm{pure}}=F_*$. Specific parameters are given in Table~\ref{tab:glued-bound-comparison}. We next compare Theorem~\ref{pure-bound} with existing pure-code bounds, explaining the relations between their parameter expressions and the extension of their scope. Write further $n-D=bM+v,\; 0\le v<b$.

\subsubsection{Extension of the Bound for Dual-Containing Classical Constructions}
By~\cite[Remark~34]{Galindo2026Locality}, when the constructed quantum code is pure, the classical distance $d_C$ in~\eqref{eq:known-galindo} can be replaced by the actual quantum distance $d$, yielding precisely~\eqref{pure-implicit}. Theorem~\ref{pure-bound} extends this parameter constraint to every pure quantum subspace satisfying Definition~\ref{def:prelim-rdelta}, without requiring a dual-containing classical construction, a stabilizer structure, or integer $k$. The distinction lies in the code class; neither result requires disjoint recovery blocks. Thus, this extension retains the original parameter expression while making it applicable to quantum subspaces that cannot be described directly by those classical constituent codes.

\paragraph{A small nonstabilizer example.}
For the pure five-qubit code $Q_5=\llbracket5,1,3\rrbracket_2$, choose a logical basis $\{|0_L\rangle,|1_L\rangle\}$ in which a weight-three representative of logical $Z$ has eigenvalues $+1$ and $-1$, respectively, and let
\begin{equation}\label{eq:pure-comparison-nonadditive}
 Q=\operatorname{span}\{
 |0_L0_L\rangle,\,
 |0_L1_L\rangle,\,
 |1_L0_L\rangle
 \}\subseteq Q_5\otimes Q_5.
\end{equation}
With the two five-qubit blocks designated as recovery blocks, $Q$ inherits local recovery from two erasures and purity for errors of weight below $3$ from the parent code. A weight-$3$ logical $Z$ on the first block acts as $\operatorname{diag}(1,1,-1)$ in this basis, so the distance is exactly $3$. Thus, $Q$ is a pure $((10,3,3))_2$ $(3,3)$-qLRC.

Since $K=3$ is not an integer power of $2$, $Q$ is not a stabilizer code and lies outside the construction class of~\cite[Theorem~31]{Galindo2026Locality}. Theorem~\ref{pure-bound}, however, applies directly. Substituting $t=2$, $b=5$, $D=2$, and $10-2=5\cdot1+3$ gives
\[
 k\le F_{\mathrm{pure}}
 =10-4-4-2\max\{0,3-3\}=2,
 \qquad\text{equivalently}\qquad K\le4.
\]
Here $k=\log_2 3$ is not an integer, demonstrating that the extension includes codes outside the original construction class.

Moreover, the right-hand side of~\eqref{pure-explicit} is always an integer. A binary code attaining this bound must therefore satisfy $K=2^{F_{\mathrm{pure}}}$. Binary codes whose dimensions are not integer powers of $2$ can thus distinguish the scopes of the bounds, but cannot attain this bound.

\subsubsection{Extension of the Disjoint-Block Singleton-like Bound}
Setting $e=D$ in~\eqref{pure-budget} gives
\[
 k\le n-2\left(
 D+t\left\lfloor\frac{n-D}{b}\right\rfloor
 \right),
\]
which is exactly the disjoint-block bound~\eqref{eq:known-pure-disjoint} from~\cite[Theorem~4]{StylianouBoche2026}. Optimizing over $e$ and using $n-D=bM+v$ gives
\[
 F_{\mathrm{pure}}
 =
 \underbrace{n-2D-2tM}_{\text{right-hand side of~\eqref{eq:known-pure-disjoint}}}
 -2\max\{0,v-r\}.
\]
Thus, the expressions coincide when $v\le r$; when $v>r$, optimization lowers the dimension exponent by $2(v-r)$.

This correction arises from the tradeoff between the global erasure budget and the number of local recovery steps. Specifically, if $v>r$, taking $e=D-(b-v)$ makes $\left\lfloor\frac{n-e}{b}\right\rfloor=M+1$. Relative to $e=D$, the global erasure budget decreases by $b-v$, while local selection adds $t$ coordinates. The total selected-set size therefore increases by $t-(b-v)=v-r$, lowering the bound by $2(v-r)$. Since $v<b=r+t$, this correction is at most $2(t-1)$; in particular, it is always zero when $\delta=2$.

Beyond this numerical improvement, Theorem~\ref{pure-bound} extends the expression to arbitrarily overlapping recovery blocks. Its proof handles overlaps through ordered selection and successive local decoupling, without requiring the designated recovery blocks to form a disjoint partition.

\subsubsection{Extension of the Explicit LP Exponent}
Suppose $n=sb$, and write $d=zb+\epsilon$ with $z\ge0$ and $1\le\epsilon\le b$. If $\epsilon=1$, then $M=s-z$ and $v=0$, so~\eqref{pure-explicit} gives
\[
 F_{\mathrm{pure}}
 =sb-2zb-2t(s-z)
 =r(s-2z)-st.
\]
If $2\le\epsilon\le b$, then $n-D=(s-z-1)b+(b-\epsilon+1)$, so $M=s-z-1$ and $v=b-\epsilon+1$. Hence
\begin{align*}
 F_{\mathrm{pure}}
 &=r(s-2z)-st
   -2(\epsilon-\delta)-2\max\{0,\delta-\epsilon\}\\
 &=r(s-2z)-st-2\max\{0,\epsilon-\delta\}.
\end{align*}
Combining the cases gives
\[
 F_{\mathrm{pure}}=
 \begin{cases}
 r(s-2z)-st,
   &1\le\epsilon\le\delta,\\
 r(s-2z)-st-2(\epsilon-\delta),
   &\delta<\epsilon\le b.
 \end{cases}
\]
This is precisely the explicit LP exponent~\eqref{eq:known-lp-explicit} from~\cite[Theorem~6]{StylianouBoche2026}. The piecewise form also shows that, for fixed $s,z$, the exponent is constant for $\epsilon\le\delta$; beyond $\delta$, each unit increase in distance lowers the exponent by two.

Thus, when $b\mid n$, the entropy argument and this explicit LP consequence give the same parameter constraint. This identifies the explicit exponent, rather than the optimum of the full linear program. Theorem~\ref{pure-bound} additionally allows arbitrary overlaps and gives an explicit bound through~\eqref{pure-explicit} for lengths with $b\nmid n$.

Theorem~\ref{pure-bound} unifies and extends several existing
pure-code parameter bounds. It extends the pure-code
constraint of Galindo et al. from dual-containing classical
constructions to arbitrary pure quantum subspaces.
It also removes the disjoint-block assumption from the
Stylianou--Boche Singleton-like bound and strengthens its
expression when $v>r$. When $b\mid n$, our bound coincides
with their explicit LP-based Singleton-like bound, thereby
extending that parameter constraint to overlapping recovery
blocks. This correspondence concerns the explicit bound,
not the full linear-programming framework.
\begin{remark}[Purity and block weight-pattern purity]
\label{pure-block-pattern}
For nontrivial codes with disjoint recovery blocks, the block weight-pattern purity of~\cite[Definition~4]{StylianouBoche2026} is equivalent to purity under our local recovery model. Indeed, purity and the local Knill--Laflamme conditions imply that every nonidentity basis error $E_J$ of weight at most $t$ on a recovery block $J$ satisfies $P_JE_JP_J=0$, since its scalar compression coefficient is $\operatorname{Tr}(\rho_JE_J)=0$. Consider a nonidentity tensor-product basis error whose block weights $i_J$ satisfy $\sum_{J:i_J\ge\delta}i_J\le D$. If it has a nonidentity factor with $i_J\le t$, the local compression identity and $Q\subseteq Q_J\otimes\mathcal H_{J^c}$ imply $PEP=0$. If it has no such factor, its total weight is at most $D$, and purity gives the same conclusion. Thus, all Shor--Laflamme coefficients for the nonzero patterns required in that work vanish. Conversely, these patterns include every error of total weight $1,\ldots,D$. Vanishing of the corresponding traces, together with the Knill--Laflamme conditions for global distance $d$, gives~\eqref{pure-definition}. The blockwise condition should therefore not be regarded as a strictly stronger purity assumption when comparing with that disjoint-block result.
\end{remark}

\subsection{Comparison at Minimal Locality}
\label{subsec:comparison-minimal}
For $\mathbb F_q$-linear stabilizer codes with $r=\delta$, Theorem~\ref{thm:minimal-locality-stabilizer} gives $\delta\mid d$. Write $b=2\delta-1$ and $\ell=d/\delta$. Then
\[
 k\le F_{\min}:=\left\lfloor\frac nb\right\rfloor-2\ell+2.
\]
Using this algebraic structure can strictly improve on bounds for general quantum subspaces. For example, when $(n,r,\delta,d)=(20,3,3,6)$,
\[
 F_{\mathrm{GG}}=6,\qquad F_{\mathrm{multi}}=4,\qquad
 F_*=3,\qquad F_{\min}=2.
\]
For every prime power $q\ge5$, the construction in Theorem~\ref{thm:minimal-locality-stabilizer} attains $k=2$. More generally, let $n=hb$, fix $\delta,\ell$, and let $h\to\infty$. Then
\[
 F_*=\frac{\delta^2}{2\delta-1}h+O(1),\qquad
 F_*-F_{\min}=\frac{(\delta-1)^2}{2\delta-1}h+O(1).
\]
The improvement over the optimized general bound therefore grows linearly with the number of blocks. On the common pure-code parameter range $n=hb$, $d=\delta$, we have $F_{\mathrm{pure}}=F_{\min}=h$. The minimal-locality theorem also applies to impure codes; the construction attaining $k=2$ above is impure.

Relative to the existing disjoint-block bound, the main advance is the derivation of the recovery structure. When $n=hb$ and $d=\delta\ell$, equation~\eqref{eq:known-multi-disjoint} already gives
\[
 k\le hb-2(\delta\ell-1)-2(\delta-1)(h-\ell)
   =h-2\ell+2,
\]
which equals $F_{\min}$. That bound assumes an equal-sized disjoint partition, whereas we prove that, after fixed pure-state coordinate factors are removed, such a partition is forced by $r=\delta$ and the linear stabilizer structure. The theorem also gives divisibility of the actual distance and attaining constructions when $b\nmid n$. The singleton-intersection restriction in~\cite[full version, Theorem~74]{Golowich2023LRC} is related; here, when $r=\delta$, all nonempty proper overlaps between the remaining blocks are excluded after the fixed pure-state coordinate factors have been removed.

Zhou and Cao~\cite{zhou2025optimalquantumrdeltalocallyrepairable,ZhouCao2025} obtain local MDS structure by decomposing optimal classical $(r,\delta)$-LRCs and characterize the induced optimal pure quantum codes under a minimal-decomposition condition. Their condition requires a classical cover consisting of $\lceil\dim(C)/r\rceil$ local groups, in which only adjacent groups may overlap. We instead start directly from $r=\delta$ and an $\mathbb F_q$-linear stabilizer structure to derive disjoint inner-code blocks and the actual-distance relation $d=\delta\ell$, without requiring purity or optimality of a classical constituent code.

\section{Constructions Attaining the Bounds}\label{sec:constructions}
We present three types of attaining constructions. A pure CSS family exhibits both a strict improvement over the general bound and unavoidable overlaps. Impure concatenated codes show that the general bound remains sharp beyond pure codes. An existing Hermitian family gives overlapping attaining examples with $d>\delta$. We use $F_{\mathrm{multi}}$, $F_*$, and $F_{\mathrm{pure}}$ for the right-hand sides of Theorem~\ref{thm:theory-bound}, Corollary~\ref{cor:theory-optimization}, and Theorem~\ref{thm:pure-overlap}, respectively.

Existing optimal pure codes include constructions based on good polynomials, Hermitian designs, and several matrix-product methods~\cite{Sharma2025Good,Li2026Hermitian,Galindo2026MPC,CaoZhou2026MPC,CaoZhou2026MP}. The examples below attain our parameter bounds, and we characterize their recovery-block overlap structures.

\subsection{A Glued GRS Family with Unavoidable Overlaps}\label{subsec:glued-grs}
The use of overlapping local check blocks appears in earlier constructions, including~\cite{zhou2025optimalquantumrdeltalocallyrepairable,ZhouCao2025}. Here we use standard GRS and CSS methods to give an explicit family, prove a lower bound on the size of \emph{every} valid recovery block, and determine the parameter ranges in which the family attains the bounds above. Shared coordinates participate directly in both sets of local checks. Support criteria for CSS local recovery can be found in~\cite{Luo2023CSS,Galindo2026Locality}.

Let $t\ge1$, $a\ge2t$, and let $q\ge a+2t$ be a prime power. Choose disjoint evaluation sets $\mathcal P=\{\alpha_1,\ldots,\alpha_a\},\; \mathcal O=\{\beta_1,\ldots,\beta_{2t}\}\subseteq\mathbb F_q$. For a set of distinct points $\mathcal T=\{z_1,\ldots,z_m\}$, define
\[
\begin{split}
V_t(\mathcal T)&=(z_j^i)_{0\le i<t,\,1\le j\le m},\\
w_j(\mathcal T)&=\left(\prod_{\ell\ne j}(z_j-z_\ell)\right)^{-1},\\
W_t(\mathcal T)&=V_t(\mathcal T)\operatorname{diag}(w_1(\mathcal T),\ldots,w_m(\mathcal T)).
\end{split}
\]
Abbreviate the corresponding matrices as $V_P,V_O,W_P,W_O$. Partition the physical coordinates into $A,B,O$ of sizes $a,a,2t$, respectively; both private sets $A,B$ use the same evaluation set $\mathcal P$. Define the $X$- and $Z$-type stabilizer matrices
\begin{equation}\label{eq:glued-checks}
H_X=\begin{pmatrix}V_P&0&V_O\\0&V_P&V_O\end{pmatrix},\qquad
H_Z=\begin{pmatrix}W_P&0&W_O\\0&W_P&W_O\end{pmatrix}.
\end{equation}
For nonprime $q$, use the usual $\mathbb F_q$-linear CSS construction and trace-form Pauli operators~\cite{Ketkar2006Nonbinary}.

\begin{theorem}[Pure attaining codes with unavoidable overlaps]\label{thm:glued-optimal}
Equation~\eqref{eq:glued-checks} defines a pure CSS code
\begin{equation}\label{eq:glued-parameters}
Q_{q,a,t}=\llbracket2a+2t,\,2a-2t,\,t+1\rrbracket_q,
\qquad(r,\delta)=(a+t,t+1).
\end{equation}
Both $\Gamma_A=A\cup O$ and $\Gamma_B=B\cup O$ are recovery blocks of size $b=a+2t$, with intersection size $2t$. Moreover, every nonempty block that corrects all sets of at most $t$ erasures within itself has size at least $b$. Any two such blocks therefore intersect in at least $2t$ coordinates, and no disjoint recovery-block partition satisfies the same local correction requirement.

For every $a\ge2t$, this code attains Theorem~\ref{thm:pure-overlap}. The optimized general bound is
\begin{equation}\label{eq:glued-bound-regimes}
F_{\mathrm{pure}}=k=2a-2t,\qquad
F_*=
\begin{cases}
a+t,&2t\le a<3t,\\
2a-2t,&a\ge3t.
\end{cases}
\end{equation}
In particular, the subfamily $a=2t$ exhibits a strict improvement of $t$ over the optimized general bound, whereas the subfamily $a\ge3t$ simultaneously attains the general and pure-code bounds.
\end{theorem}
\begin{proof}
\emph{Commutation and dimension.}
Lagrange interpolation gives
\[
\sum_{j=1}^{m}w_j(\mathcal T)z_j^u=0
\qquad(0\le u\le m-2).
\]
Since both $|\mathcal P|=a$ and $|\mathcal O|=2t$ are at least $2t$, we have $V_PW_P^{\mathrm T}=V_OW_O^{\mathrm T}=0$, and hence $H_XH_Z^{\mathrm T}=0$. Restricting a row relation to $A$ and $B$ separately shows that each matrix has rank $2t$. Thus $n=2a+2t$ and $k=n-4t$.

\emph{Local supports and recovery.}
Taking a partial trace of a normalized stabilizer-code projector retains exactly the stabilizers supported on the retained coordinates. For a CSS code, the two local check spaces are therefore the shortenings of the two global check spaces. A check supported on $\Gamma_A$ vanishes on $B$. Since $V_P$ and $W_P$ have full row rank, its coefficients in the second group of rows must vanish. The local support code therefore has check matrices exactly
\begin{equation}\label{eq:glued-local-checks}
M_X=[V_P\ V_O],\qquad M_Z=[W_P\ W_O].
\end{equation}
Both are GRS matrices of length $b$ and dimension $t$, and any $t$ columns are linearly independent. No nonidentity Pauli operator of weight at most $t$ commutes with all local checks. The local code has dimension parameter $b-2t=a$ and distance at least $t+1$, so the quantum Singleton bound gives the exact parameters $Q_{\Gamma_A}=\llbracket a+2t,a,t+1\rrbracket_q$. The same holds on $\Gamma_B$. By the local support criterion, the erasure recovery channel of this local code uses only its block and restores the global encoded state together with its correlations with an external reference system.

\emph{Exact global distance and purity.}
If $x=(x_A,x_B,x_O)\in\ker H_Z$ has weight at most $t$, the first group of equations and column independence of $M_Z$ imply $x_A=x_O=0$; the second group then gives $x_B=0$. The same argument applies to $\ker H_X$. Thus the global distance is at least $t+1$. Choose any $t+1$ coordinates in $A$. The restriction of $W_P$ to these columns has a one-dimensional kernel, and every component of a nonzero kernel vector is nonzero. Extending by zeros gives $x\in\ker H_Z$ of weight $t+1$. This vector is not in the row space of $H_X$: its zero $B$ component first forces the second group of coefficients to vanish, and its zero $O$ component then forces the first group to vanish. Thus $X(x)$ is a nontrivial logical operator and $d=t+1$.

Every $X$ check can be represented by two polynomials of degree less than $t$ as
\[
\bigl((f(\alpha_i))_i,(g(\alpha_i))_i,((f+g)(\beta_j))_j\bigr).
\]
If exactly one of $f,g$ is nonzero, its weight is at least $a+t+1$. If both are nonzero, the two private parts alone have total weight at least $2(a-t+1)$. Nonzero column multipliers give the same estimate for $Z$ checks. The support of a mixed check is the union of the supports of its two components, so every nonidentity stabilizer has weight at least
\begin{equation}\label{eq:glued-stabilizer-weight}
\min\{a+t+1,\,2(a-t+1)\}\ge2t+2>t+1.
\end{equation}
This proves purity for all $a\ge2t$.

\emph{A support lower bound for all recovery blocks.}
Let $U$ be any $t$-dimensional subspace of the $X$-check space, represented by polynomial pairs $(f,g)$. Let its projections onto $f,g,f+g$ have dimensions $u,v,w$, respectively; then $u+v\ge t$. Every $e$-dimensional subspace of polynomials of degree less than $t$ has at most $t-e$ common zeros. Otherwise, dividing out the corresponding common linear factors leaves a polynomial space of dimension less than $e$, a contradiction. If $u=0$ or $v=0$, the other polynomial ranges over all polynomials of degree less than $t$, and the union of the supports in $U$ has size $b$. If $u,v>0$ and $w=0$, then $g=-f$, so $u=v=t$ and the support size is $2a\ge b$. Finally, if $u,v,w>0$, the three coordinate parts give
\[
|\operatorname{supp}U|
\ge(a-t+u)+(a-t+v)+(t+w)
\ge2a+w\ge b.
\]
All column multipliers are nonzero, so the same argument applies to the $Z$-check space.

Now let $\Gamma$ be a nonempty block that corrects any set of at most $t$ erasures within itself. If $|\Gamma|\le t$, equation~\eqref{eq:glued-stabilizer-weight} makes its local support code the full space, which cannot correct even one erasure. Hence $|\Gamma|>t$. If the $Z$-check space supported on $\Gamma$ had dimension below $t$, the restriction to any $t$ columns would have a nonzero kernel vector. The corresponding $X$ operator would have weight at most $t$ and commute with all local checks. By~\eqref{eq:glued-stabilizer-weight}, it could not be a local stabilizer, contradicting local erasure correctability. The shortened $Z$-check space therefore contains a $t$-dimensional subspace, and the support bound above implies $|\Gamma|\ge b$. Consequently,
\[
|\Gamma\cap\Gamma'|\ge2b-n=2t.
\]

\emph{Attainment and comparison.}
Here $D=d-1=t$, $b=a+2t$, and $n-D=b+(a-t)$. Since the remainder $a-t$ is less than $r=a+t$, Theorem~\ref{thm:pure-overlap} gives $F_{\mathrm{pure}}=n-2t-2t=2a-2t$. For the general bound, $L=n-2d+2=2a$, and the largest first-round local erasure count is $t$. Hence
\[
F_*=\min_{0\le j\le t}\left\{2a-j-t\left\lfloor\frac{2a-j}{a+2t}\right\rfloor\right\},
\qquad j\in\mathbb Z.
\]
If $a\ge3t$, the floor is always $1$, giving $F_*=2a-2t$. If $2t\le a<3t$, the floor equals $1$ exactly when $j\le a-2t$. The minima on the two intervals are $a+t$ and $2a-t$, respectively. Since $a\ge2t$, the smaller is $a+t$. This proves~\eqref{eq:glued-bound-regimes}.
\end{proof}

The local recovery channel
$\operatorname{Rec}_{\Gamma_A,E}:\mathcal B(\mathcal H_{\Gamma_A\setminus E})
\to\mathcal B(\mathcal H_{\Gamma_A})$, for $|E|\le t$, also has a direct
implementation. On input $\sigma$ on the surviving registers
$\Gamma_A\setminus E$, prepare new registers $E$ in the maximally mixed
state and measure the local stabilizers. When $\sigma$ is obtained by
tracing out $E$ from a local code state, the prepared state equals the
uniform Pauli twirl of that code state on $E$.
The full-column-rank matrices $(M_Z)_E$ and $(M_X)_E$ determine the two
error vectors uniquely from the syndrome, so the corresponding inverse
Pauli restores the state. For any syndrome incompatible with an error
supported on $E$, choose the identity correction. Over a nonprime field,
a complete set of trace syndromes suffices. Averaging over all measurement outcomes
defines a CPTP map on the entire input space. Its output registers are
$\Gamma_A$; it does not access $B$ and preserves correlations with an
external reference system. The endpoint subfamily is
\begin{equation}\label{eq:glued-endpoint-family}
\llbracket6t,2t,t+1\rrbracket_q,
\qquad(r,\delta)=(3t,t+1),\quad q\ge4t.
\end{equation}
It satisfies $F_*=3t$ and $F_{\mathrm{pure}}=2t$, and any two valid recovery blocks intersect in at least $2t$ coordinates. Table~\ref{tab:glued-bound-comparison} gives two adjacent examples illustrating strict improvement and simultaneous attainment. The rows use $a=5$ and $a=6$, respectively, with $t=2$ in both. Over $\mathbb F_{11}$, take $\mathcal P=\{0,\ldots,a-1\}$ and $\mathcal O=\{a,\ldots,a+3\}$. Every valid recovery block has at least $a+4$ coordinates, and any two intersect in at least four coordinates.
\begin{table}[t]
\centering
\caption{Two roles of the same GRS construction.}\label{tab:glued-bound-comparison}
\begin{tabular}{ccrrr}
\toprule
Code & $(r,\delta)$ & $F_{\mathrm{multi}}$ & $F_*$ & $F_{\mathrm{pure}}$\\
\midrule
$\llbracket14,6,3\rrbracket_{11}$ & $(7,3)$ & $8$ & $7$ & $6$\\
$\llbracket16,8,3\rrbracket_{11}$ & $(8,3)$ & $8$ & $8$ & $8$\\
\bottomrule
\end{tabular}
\end{table}
An extension to multiple blocks sharing a common core is given in Appendix~\ref{app:sunflower}.

\subsection{Impure Attaining Families from Standard CSS Concatenation}
The following families use nested Reed--Solomon CSS codes~\cite{Ketkar2006Nonbinary} and standard quantum concatenation~\cite{Hamada_2008}. They show that impure codes can attain the general bound; the construction methods themselves are established in the literature.
\begin{proposition}[Impure attaining families]\label{prop:impure-attaining-families}
For every $\delta\ge2$ and prime power $q\ge2\delta-1$, there exists an impure CSS $(\delta,\delta)$-qLRC
\[
Q^{(3)}_{q,\delta}=\llbracket6\delta-3,1,2\delta\rrbracket_q.
\]
For every prime power $q\ge4$, there also exists an impure CSS $(2,2)$-qLRC $\llbracket12,2,4\rrbracket_q$. These families have three and four disjoint designated recovery blocks, respectively, and attain Theorem~\ref{thm:theory-bound} and Corollary~\ref{cor:theory-optimization}.
\end{proposition}
\begin{proof}
Let $\operatorname{RS}_q(\ell,u)$ denote the code obtained by evaluating polynomials of degree less than $u$ at $\ell$ distinct field elements. The nested pair
\[
D_{\rm in}=\operatorname{RS}_q(2\delta-1,\delta-1)
\subset C_{\rm in}=\operatorname{RS}_q(2\delta-1,\delta)
\]
defines a CSS inner code with stabilizers $X(D_{\rm in})$ and $Z(C_{\rm in}^{\perp})$. The codes $C_{\rm in}$ and $D_{\rm in}^{\perp}$ governing logical distance both have distance $\delta$, whereas both stabilizer components have distance $\delta+1$. The inner code therefore has exact parameters $\llbracket2\delta-1,1,\delta\rrbracket_q$. For $h=3$ or $4$ and $q\ge h$, the nested outer pair $\operatorname{RS}_q(h,1)\subset\operatorname{RS}_q(h,h-1)$ gives $\llbracket h,h-2,2\rrbracket_q$. Concatenation yields length $h(2\delta-1)$, dimension parameter $h-2$, and distance at least $2\delta$.

To prove that the distance is exactly $2\delta$, choose a weight-two vector in $C_{\rm out}\setminus D_{\rm out}$, representing a nontrivial outer $X$ logical operator. Every nonzero coset of the one-dimensional quotient $C_{\rm in}/D_{\rm in}$ has a representative of weight $\delta$: nonzero scalar multiples of a minimum-weight vector of $C_{\rm in}$ cover all nonzero quotient elements. Lifting the two nonzero components of the outer vector to these representatives produces a nontrivial physical logical operator of weight $2\delta$. Each inner block locally corrects $\delta-1$ erasures, and its weight-$\delta+1$ stabilizers remain after concatenation. Since $\delta+1<2\delta$, the concatenated code is impure.

For $h=3$, the first-round local deduction in Theorem~\ref{thm:theory-bound} is $2(\delta-1)$, and the remaining second-round budget is $1$, giving $k\le1$. For $h=4$, $\delta=2$, the two deductions are $3$ and $1$, giving $k\le2$. Both bounds are attained. The optimized bound is no larger than the general bound and also applies to these existing codes, so it too equals their dimension parameter.
\end{proof}
When $q=3^m$ and $\delta=(q+1)/2$, the first family reproduces the parameters of the impure family in~\cite[Proposition~6]{Galindo2026Impure}; we do not claim equivalence of the codes produced by the two constructions. When $\delta>2$, this family exhibits a strict improvement over the single-erasure benchmark. The second family shows that impure codes can also attain the bound when $k>1$ and both local deductions are nonzero.

\subsection{An Existing Hermitian Family with \texorpdfstring{$d>\delta$}{d > delta}}\label{subsec:existing-overlap-attainment}
Unlike the glued GRS family above, the following overlapping codes have global distance strictly greater than their local distance. The construction is due to Zhou and Cao~\cite{zhou2025optimalquantumrdeltalocallyrepairable,ZhouCao2025}; here we verify attainment of the stated bounds and compatibility with the recovery-channel model.
\begin{proposition}[Hermitian attaining families]
\label{prop:existing-overlap-families}
Let $s\ge2$, $c\in\{1,2\}$, and let $q\ge3s+2c+1$ be a prime power.
There exists a pure code
\begin{equation}\label{eq:existing-overlap-parameters}
\llbracket3q-3,\,3q-3-4s-2c,\,s+c+1\rrbracket_q,
\qquad(r,\delta)=(2q-s-2,s+1).
\end{equation}
Its two designated recovery blocks each have $2(q-1)$ coordinates
and intersect in $q-1$ coordinates. The code attains
$F_{\mathrm{multi}}$, $F_*$, and $F_{\mathrm{pure}}$.
\end{proposition}
\begin{proof}
We specialize the construction of Zhou and Cao
~\cite[arXiv version, Theorems~5.4 and~5.6]{zhou2025optimalquantumrdeltalocallyrepairable,ZhouCao2025}.
Set $G=\mathbb F_q^\times\subset\mathbb F_{q^2}$ and $m=q-1$,
and choose a primitive element $\gamma\in\mathbb F_{q^2}$.
Partition the coordinates as $A\sqcup O\sqcup B$, each part of size $m$,
with evaluation sets $X=\gamma G$, $Y=G$, and
$Z=\gamma G$ for $c=1$ or $Z=\gamma^2G$ for $c=2$.
Repeated evaluation points in different parts represent distinct
physical coordinates. Write
$V_s(T)=(z^i)_{1\le i\le s,\,z\in T}$ and
$W_{s,c}(T)=(z^{s+i})_{1\le i\le c,\,z\in T}$.
The classical parity-check matrix is
\begin{equation}\label{eq:existing-overlap-H}
H=\begin{pmatrix}
V_s(X)&V_s(Y)&0\\
0&V_s(Y)&V_s(Z)\\
W_{s,c}(X)&W_{s,c}(Y)&W_{s,c}(Z)
\end{pmatrix},
\qquad C=\ker H.
\end{equation}
This corresponds to one pair of overlapping blocks with overlap
parameter $v=m$. The conditions
$s\le\lfloor v/2\rfloor-c-1$ and
$v\le q+v-s-c-2$ follow from $s\ge2$ and $q\ge3s+2c+1$.
The cited construction therefore gives a pure Hermitian quantum code
with the parameters in~\eqref{eq:existing-overlap-parameters}.
We next verify local recovery in our channel model.
With the convention $C=\ker H$, the Hermitian stabilizer
label space is $C^{\perp_h}=\operatorname{row}(H^{(q)})$,
where $H^{(q)}$ denotes entrywise $q$th powers.
Taking entrywise $q$th powers, a stabilizer label supported
on $\Gamma_1=A\cup O$ corresponds to a row combination of
$H$ that vanishes on $B$.
The corresponding polynomial has degree at most $s+c<m$,
so vanishing at all points of $Z$ forces the coefficients
of the global rows and the second group of local rows to vanish.
Thus the shortened stabilizer label space is the
$\mathbb F_{q^2}$-span of the $q$th powers of the first
$s$ local rows.
These conjugated Vandermonde rows define a pure local MDS
code $\llbracket2m,2m-2s,s+1\rrbracket_q$.
The same argument applies to $\Gamma_2=O\cup B$.
Lemma~\ref{lem:theory-support} supplies recovery channels acting only
within these blocks. Since $|\Gamma_i|=2m=r+\delta-1$ and
$\Gamma_1\cap\Gamma_2=O$, the stated locality and overlap follow.

Finally, $t=s$, $D=s+c$, $b=2m$, and $L=3m-2s-2c$.
The assumption $m\ge3s+2c$ gives
\[
\left\lfloor\frac{n-D}{b}\right\rfloor
=\left\lfloor\frac{L-s}{b}\right\rfloor=1,
\]
so $F_{\mathrm{multi}}=L-2s=k$.
For the pure-code bound,
$n-D=b+(m-s-c)$, where $0\le m-s-c<r=2m-s$.
Hence $F_{\mathrm{pure}}=L-2s=k$ as well.
Since $k\le F_*\le F_{\mathrm{multi}}$, the optimized general bound
is also attained.
\end{proof}
For example, $s=2$, $c=2$, $q=11$ give $\llbracket30,18,5\rrbracket_{11}$ with local parameters $(r,\delta)=(18,3)$ and two recovery blocks intersecting in ten coordinates~\cite[arXiv version, Example~5.8]{zhou2025optimalquantumrdeltalocallyrepairable,ZhouCao2025}. Here $d=\delta+2$. For these existing families, we assert only that the designated recovery blocks overlap. The stronger conclusion excluding every alternative disjoint recovery partition was proved above for the glued GRS family.

\begin{remark}[The quantum MDS boundary]\label{prop:mds-overlap-boundary}
When $q\ge n\ge2d-1$, the standard nested Reed--Solomon CSS construction also gives a pure $\llbracket n,n-2d+2,d\rrbracket_q$ code. For $2\le\delta<d$, shortening its check spaces shows that every $(n-d+\delta)$-element subset supports a local $\llbracket n-d+\delta,n-d-\delta+2,\delta\rrbracket_q$ MDS code, giving an overlapping $(n-d+1,\delta)$-qLRC. Both local deduction terms in the general bound vanish, so attainment here is ordinary quantum Singleton attainment.
\end{remark}

\section{Conclusion}\label{sec:conclusion}

We established Singleton-like bounds for general $(r,\delta)$-qLRCs
without assuming purity, a stabilizer structure, or disjoint recovery
blocks. These bounds address the gap between general single-erasure
bounds and multiple-erasure bounds requiring additional structural
assumptions. Integer optimization yields an explicit remainder
correction, while purity gives a stronger bound that extends several
existing parameter expressions to arbitrary pure quantum subspaces.
We also derived structural constraints under equality, proved a
decomposition theorem for $\mathbb F_q$-linear stabilizer codes when
$r=\delta$, and analyzed families attaining the bounds with different
purity and overlap properties.

Future work includes developing alphabet-dependent bounds without
partition assumptions, determining the full range of parameters for
which the bounds are attainable, and investigating which structural
conclusions extend beyond $\mathbb F_q$-linear stabilizer codes.

\appendices
\section{Proof of the Minimal-Locality Structure Theorem}\label{app:minimal-locality-proof}
\begin{proof}
We first examine the recovery-block structure. By the local support equivalence, each local support corrects any set of at most $t$ erasures within its block, and its dimension is a power of $q$. If this dimension is greater than $1$, the local quantum Singleton bound and $|\Gamma|\le b=2t+1$ give
\[
q\le\dim Q_\Gamma\le q^{|\Gamma|-2t}\le q.
\]
Thus $|\Gamma|=b$, and the local support has parameters $\llbracket b,1,t+1\rrbracket_q$. Denote its $2t$-dimensional local stabilizer space by $S_\Gamma=S[\Gamma]$.

The MDS entropy formula~\eqref{eq:local-mds-entropy-profile} gives $S_\Gamma[T]=0$ for $|T|\le t+1$. Correctability also gives $S_\Gamma^{\perp_s}\cap V_T=S_\Gamma\cap V_T=0$ for $|T|\le t$. Counting dimensions on $t+2$ and $t+1$ coordinates, respectively, yields
{
\begin{equation}\label{eq:local-mds-stabilizer-facts}
d(S_\Gamma)=t+2,\qquad d(S_\Gamma^{\perp_s})=t+1.
\end{equation}
}
Here the dual is taken only within $\Gamma$. These are standard quantum MDS stabilizer properties~\cite{Ketkar2006Nonbinary}; the entropy argument above proves the case needed here directly. The projection of $S_\Gamma$ onto any $u\le t$ coordinates is surjective, since its symplectic annihilator $S_\Gamma^{\perp_s}\cap V_T$ is zero.

Consider distinct blocks $\Gamma,\Delta$ whose local supports both have dimension $q$, and write $s=|\Gamma\cap\Delta|$. If $1\le s\le t$, both local stabilizers project surjectively onto the intersection, so their symplectic inner product cannot vanish for every pair of vectors. Their supports are disjoint outside the intersection, contradicting global stabilizer self-orthogonality. If $t<s<b$, then $u=b-s\le t$, and surjectivity onto $\Gamma\setminus\Delta$ gives
\[
\dim(S_\Gamma\cap S_\Delta)
\le\dim(S_\Gamma\cap V_{\Gamma\cap\Delta})
=2t-2u=2s-2t-2.
\]
Consequently,
\[
6t-2s+2\le\dim(S_\Gamma+S_\Delta)
\le|\Gamma\cup\Delta|=4t+2-s,
\]
where the second inequality uses the maximum dimension of a symplectically self-orthogonal space on the union. Thus $s\ge2t$, and the only possible proper overlap is $s=2t=b-1$. In that case, $S_\Gamma+S_\Delta$ has the maximum self-orthogonal dimension on the union, whose reduced state is therefore pure.

We now justify the removal of fixed pure factors. Let $\rho=P/q^k$ and choose a coordinate set $U$ of maximum cardinality such that $\rho_U$ is pure; $U$ may be empty. Then $\rho=|\phi\rangle\langle\phi|_U\otimes\rho_{U^c}$. Since $\dim S[U]=|U|$, symplectic elimination shows that the remaining factor is still an $\mathbb F_q$-linear stabilizer code. A one-dimensional local support is itself a pure marginal, so its entire block must lie in $U$. Here we use the fact that, if the marginals on two coordinate sets are both pure, then the marginal on their union is also pure. This follows by applying the pure-marginal factorization to their intersection and differences.

A block whose local support has dimension $q$ cannot partially intersect $U$. Its normalized projector satisfies the entropy formula~\eqref{eq:local-mds-entropy-profile}. If $\Gamma$ met both $U$ and $U^c$, the global factorization would make its marginal a product across this nontrivial partition. Yet~\eqref{eq:local-mds-entropy-profile} gives $S(\Gamma)=1$, whereas the entropies of the two parts sum to at least $3$, a contradiction. Thus every coordinate in $U^c$ belongs to a recovery block contained entirely in $U^c$, of length $b$, and with local support dimension $q$. Two distinct such blocks cannot overlap, because the preceding overlap analysis would produce an additional pure marginal, contradicting maximality of $U$. These blocks therefore partition $U^c$ into $h$ disjoint blocks.

Denote the blocks by $\Gamma_1,\ldots,\Gamma_h$ and the stabilizer of $Q'$ by $S'$. The direct sum $S_0=\bigoplus_jS_{\Gamma_j}$ is contained in $S'$, and $S'\subseteq S_0^{\perp_s}$. The symplectic quotient
\[
S_0^{\perp_s}/S_0
\simeq\bigoplus_{j=1}^h
\left(S_{\Gamma_j}^{\perp_s}/S_{\Gamma_j}\right)
\simeq\mathbb F_q^{2h}
\]
identifies $S'/S_0$ with a symplectically self-orthogonal space of dimension $h-k$, giving the asserted linear outer stabilizer code.

It remains to establish the exact distance, rather than only the product lower bound. Fix any $E\subseteq\Gamma_j$ of size $|E|=t+1$. The projection of $S_{\Gamma_j}$ onto $E$ is injective, since a vector in its kernel would be supported on at most $t$ coordinates. Hence
\[
\dim\bigl(S_{\Gamma_j}^{\perp_s}\cap V_E\bigr)=2|E|-2t=2,
\qquad S_{\Gamma_j}\cap V_E=0.
\]
This two-dimensional space is isomorphic to the logical quotient $S_{\Gamma_j}^{\perp_s}/S_{\Gamma_j}$. Every nonzero logical Pauli coset therefore has a representative of weight exactly $t+1=\delta$. A physical logical operator of $Q'$ maps to an outer logical operator, and each nonzero outer coordinate requires physical weight at least $\delta$. Conversely, lifting a minimum-weight outer logical operator block by block to these representatives gives a physical logical operator of weight exactly $\delta d_{\mathrm{out}}$. The stabilizer on $U$ equals its symplectic dual, so the component on $U$ of any normalizer element can be removed by multiplication by a stabilizer without changing its logical class. The fixed pure factor therefore does not affect the logical distance, and the original code also satisfies $d=\delta d_{\mathrm{out}}$. Applying the quantum Singleton bound to the outer code proves~\eqref{eq:minimal-locality-sharp}.

To prove sharpness, use the standard nested Reed--Solomon CSS constructions~\cite{Ketkar2006Nonbinary,Hamada_2008}:
\[
\operatorname{RS}_q(b,t)\subset\operatorname{RS}_q(b,t+1),
\qquad
\operatorname{RS}_q(h,d_{\mathrm{out}}-1)\subset\operatorname{RS}_q(h,h-d_{\mathrm{out}}+1),
\]
which give an inner code $\llbracket b,1,\delta\rrbracket_q$ and an outer code $\llbracket h,h-2d_{\mathrm{out}}+2,d_{\mathrm{out}}\rrbracket_q$, respectively. The classical codes governing logical distance have distances $\delta$ and $d_{\mathrm{out}}$, while the corresponding stabilizer components have strictly larger distances. For $d_{\mathrm{out}}=1$, take the full space as the outer code. The field-size assumption ensures enough distinct evaluation points. Concatenate the two codes and append $z$ qudits in a fixed product state. The logical-coset argument above gives exact distance $\delta d_{\mathrm{out}}$; the inner blocks and singleton blocks on the appended coordinates give the required locality.
\end{proof}

\section{Multiple Local Blocks Sharing a Common Core}\label{app:sunflower}
The matrices in~\eqref{eq:glued-checks} extend to any prescribed number of blocks. We record a parameter range attaining both types of bounds. The overlap property here concerns only the designated recovery blocks; it does not assert that every pair of alternative recovery blocks must intersect.
\begin{corollary}[CSS families with a shared core]
\label{prop:glued-sunflower}
Let $h\ge3$, $t\ge1$, and $a\ge3(h-1)t$, and let $q\ge a+2t$
be a prime power. There exists a pure CSS code
\[
\llbracket ha+2t,\,ha-2(h-1)t,\,t+1\rrbracket_q,
\qquad(r,\delta)=(a+t,t+1),
\]
whose $h$ designated recovery blocks each contain $a+2t$ coordinates,
with every pair intersecting in the same common set of size $2t$.
This code attains $F_{\mathrm{multi}}$, $F_*$, and $F_{\mathrm{pure}}$.
\end{corollary}
\begin{proof}
Take $h$ pairwise disjoint private sets $A_1,\ldots,A_h$, each of size
$a$, and a common set $O$ of size $2t$. Using the notation of
Section~\ref{subsec:glued-grs}, set
\[
\begin{aligned}
H_X&=[\operatorname{diag}(V_P,\ldots,V_P)
      \mid \mathbf 1_h\otimes V_O],\\
H_Z&=[\operatorname{diag}(W_P,\ldots,W_P)
      \mid \mathbf 1_h\otimes W_O].
\end{aligned}
\]
The interpolation identity gives
$V_PW_P^{\mathrm T}=V_OW_O^{\mathrm T}=0$, so
$H_XH_Z^{\mathrm T}=0$. Restriction to the private sets shows that
each matrix has rank $ht$. Thus the CSS code has length $n=ha+2t$
and dimension parameter $k=n-2ht=ha-2(h-1)t$.

For $\Gamma_i=A_i\cup O$, vanishing on every other private set
forces all row coefficients outside the $i$th group to be zero.
The shortened check matrices are therefore $[V_P\ V_O]$ and
$[W_P\ W_O]$, giving the same local MDS code as in
Theorem~\ref{thm:glued-optimal}.
Hence each $\Gamma_i$ corrects any set of at most $t$ erasures
within itself. These blocks cover all coordinates and satisfy
$|\Gamma_i|=a+2t=r+\delta-1$ and
$\Gamma_i\cap\Gamma_j=O$ for $i\ne j$.

The column-independence argument in that theorem excludes nonzero
vectors of weight at most $t$ from both $\ker H_X$ and $\ker H_Z$,
so the global distance is at least $t+1$.
Conversely, any $t+1$ columns of $W_P$ support a nonzero kernel vector.
Extending it by zeros gives an $X$ logical operator on one private
set: it cannot lie in the row space of $H_X$, since vanishing on the
other private sets and on $O$ would force all row coefficients to
vanish. Thus $d=t+1$.
Every nonidentity stabilizer has a nonzero polynomial component on
some private set, giving weight at least $a-t+1>t+1$ and proving purity.

For attainment, write $D=t$, $L=n-2D=ha$, and $b=a+2t$.
The two relevant divisions are
\[
\begin{aligned}
n-D&=(h-1)b+v_1,
&v_1&=a-(2h-3)t,\\
L-(h-1)t&=(h-1)b+v_2,
&v_2&=a-3(h-1)t.
\end{aligned}
\]
The assumption $a\ge3(h-1)t$ ensures
$0\le v_2\le v_1<r=a+t<b$.
Consequently, both floor terms in the general two-round bound equal
$h-1$, and
$F_{\mathrm{multi}}=L-2(h-1)t=k$.
Since $v_1<r$, the remainder correction in the pure-code bound
vanishes, giving
$F_{\mathrm{pure}}=n-2t-2(h-1)t=k$.
Finally, $k\le F_*\le F_{\mathrm{multi}}$ implies $F_*=k$.
\end{proof}

\section*{Acknowledgment}
The authors used GPT-6 to generate the schematic illustrations
in Figs.~1 and~2. GPT-6 also helped identify and correct an
error in an earlier version of the proof of Lemma~14,
assisted with completing the proof of Theorem~28, and
assisted in developing the constructions in
Section~\ref{sec:constructions}.
The authors reviewed and verified all AI-assisted content
and take full responsibility for the manuscript.
\bibliographystyle{IEEEtran}
\bibliography{references}
\end{document}